\documentclass[journal]{IEEEtran}
\usepackage{amsmath,amsfonts}
\usepackage{algorithmic}
\usepackage{algorithm}
\usepackage{array}
\usepackage[caption=false,font=normalsize,labelfont=sf,textfont=sf]{subfig}
\usepackage{textcomp}
\usepackage{stfloats}
\usepackage{url}
\usepackage{verbatim}
\usepackage{graphicx}
\usepackage{cite}
\usepackage[normalem]{ulem}
\usepackage{algorithmic}
\usepackage{graphicx}
\usepackage{textcomp}
\usepackage{wrapfig}

\usepackage{tikz}
\usepackage{tikzit}

\graphicspath{{figures/}}

\tikzstyle{style_0}=[fill=white, draw=black, shape=circle, minimum size=1cm]
\tikzstyle{style_1}=[fill=white, draw=black, shape=circle, minimum size=1.2cm]
\tikzstyle{style_2}=[fill=white, draw=black, shape=rectangle, thick=0.1cm, tikzit shape=rectangle, fill opacity=0]
\tikzstyle{black_fill}=[fill=black, draw=black, shape=circle, minimum size=0.01cm]

\tikzstyle{edge_0}=[->]

\usepackage{algorithmic}
\usepackage{algorithm}
\usepackage{array}
\usepackage[caption=false]{subfig}
\usepackage{textcomp}
\usepackage{stfloats}
\usepackage{url}
\usepackage{verbatim}
\usepackage[utf8]{inputenc}
\usepackage{pgfplots}
\usepgfplotslibrary{groupplots,dateplot, polar}
\usetikzlibrary{patterns,shapes.arrows,spy,matrix,positioning}
\pgfplotsset{compat=newest}
\usepackage{pgfplotstable}
\usepackage{xcolor}
\usepackage{multirow}
\usepackage{hyperref}

\begin{document}

\bstctlcite{MyBSTcontrol}

\title{MMWiLoc: A Multi-Sensor Dataset and Robust Device-Free Localization Method Using Commercial Off-The-Shelf Millimeter Wave Wi-Fi Devices}

\author{Wenbo Ding, Yang Li,~\IEEEmembership{Member, IEEE}, Dongsheng Wang, Bin Zhao, Yunrong Zhu, Yibo Zhang, Yumeng Miao  
\thanks{ This work was supported by the National Natural Science Foundation of China (NSFC) under Grant 62071156. (Corresponding author: Yang Li.)}
\thanks{ The authors are with the Research Institute of Electronic Engineering, Harbin Institute of Technology, Harbin, China (e-mail: li.yang@hit.edu.cn)}}

\markboth{Journal of \LaTeX\ Class Files,~Vol.~14, No.~8, August~2021}%
{Shell \MakeLowercase{\textit{et al.}}: A Sample Article Using IEEEtran.cls for IEEE Journals}


\maketitle

\begin{abstract}
	Device-free Wi-Fi sensing has numerous benefits in practical settings, as it eliminates the requirement for dedicated sensing devices and can be accomplished using current low-cost Wi-Fi devices. With the development of Wi-Fi standards, millimeter wave Wi-Fi devices with 60GHz operating frequency and up to 4GHz bandwidth have become commercially available. Although millimeter wave Wi-Fi presents great promise for Device-Free Wi-Fi sensing with increased bandwidth and beam-forming ability, there still lacks a method for localization using millimeter wave Wi-Fi. Here, we present two major contributions: First, we provide a comprehensive multi-sensor dataset that synchronously captures human movement data from millimeter wave Wi-Fi, 2.4GHz Wi-Fi, and millimeter wave radar sensors. This dataset enables direct performance comparisons across different sensing modalities and facilitates reproducible researches in indoor localization. Second, we introduce MMWiLoc, a novel localization method that achieves centimeter-level precision with low computational cost. MMWiLoc incorporates two components: beam pattern calibration using Expectation Maximization and target localization through Multi-Scale Compression Sensing. The system processes beam Signal-to-Noise Ratio (beamSNR) information from the beam-forming process to determine target Angle of Arrival (AoA), which is then fused across devices for localization. Our extensive evaluation demonstrates that MMWiLoc achieves centimeter-level precision, outperforming 2.4GHz Wi-Fi systems while maintaining competitive performance with high-precision radar systems. The dataset and examples processing code will be released after this paper is accepted at \url{https://github.com/wowoyoho/MMWiLoc}.
\end{abstract}

\begin{IEEEkeywords}
  Wi-Fi Sensing, Indoor Localization, Millimeter Wave Wi-Fi
\end{IEEEkeywords}

\section{Introduction}
\label{section1}

Indoor human localization has become essential for numerous applications including healthcare monitoring, emergency response, and smart building management. Over the years, numerous technologies have been developed, each presenting distinct trade-offs in real-world deployments. For example, vision-based systems \cite{mautzSurveyOpticalIndoor2011} and infrared sensors \cite{gorostizaInfraredSensorSystem2011} can provide high precision; however, they suffer from privacy concerns and high infrastructure costs.

In contrast, radio frequency (RF) sensing approaches offer promising non-intrusive alternatives. millimeter wave radar \cite{bartolettiSensorRadarNetworks2014} and Wi-Fi-based methods \cite{luoVisionTransformersHuman2024} have emerged as viable solutions. Among these, device-free Wi-Fi localization has attracted considerable attention due to its cost-effectiveness and its ability to leverage existing infrastructure \cite{liIndoTrackDeviceFreeIndoor2017}, although achieving reliable performance in complex indoor environments remains challenging.

Traditional device-free Wi-Fi localization typically estimates human position or displacement by extracting features such as Angle of Arrival (AoA) \cite{liDynamicMUSICAccurateDevicefree2016}, Time of Flight (ToF) \cite{tanMultiTrackMultiUserTracking2019}, or Doppler Frequency Shift (DFS) \cite{qianWidar2PassiveHuman2018} from Channel State Information (CSI). While conventional 2.4GHz and 5 GHz Wi-Fi systems (802.11ac/ax) have demonstrated reasonable accuracy, the emergence of millimeter wave Wi-Fi (802.11ad/ay) technology \cite{nitsche_ieee_2014} offers the potential for substantial improvements in localization precision. This is primarily due to its wider bandwidth and enhanced beam-forming capabilities, which enable the use of beam-forming techniques to systematically scan spatial sectors and collect beam Signal-to-Noise Ratio (beamSNR) information.

Despite these advantages, current millimeter wave Wi-Fi devices—typically employing a single-carrier architecture—face inherent challenges in producing ideal directional beam patterns across all sectors \cite{steinmetzer_compressive_2017}. As a result, the actual antenna radiation patterns often deviate from theoretical models \cite{bielsa_indoor_2018}, rendering direct pattern-based localization approaches unreliable. To overcome such limitations, researchers have explored deep learning solutions with commercial off-the-shelf (COTS) devices, but these approaches suffer from limited generalizability, overfitting issues, and require substantial retraining when deployed in new environments \cite{vaca-rubio_object_2024}.

Another significant hurdle is the lack of comprehensive, synchronized multi-sensor datasets. Existing datasets typically focus on a single modality or lack synchronized multi-sensor data\cite{yang_mm-fi_2023, jiang_towards_2020, liao_wisense_2024, thanh_public_2024, huang_wimans_2024}, impeding the validation of new algorithms and benchmarking performance across different technologies. This limitation is especially acute when developing and evaluating millimeter wave Wi-Fi sensing systems, where comparing performance with established technologies like radar and conventional Wi-Fi is crucial.


To address these challenges, we propose MMWiLoc, a novel localization system with three key components: (1) a comprehensive multi-sensor platform integrating synchronized radar, 2.4GHz Wi-Fi, and millimeter wave Wi-Fi data collection, (2) an enhanced Expectation Maximization algorithm for beam-forming pattern refinement, and (3) a Multi-Scale Compressive Sensing algorithm for accurate AoA estimation and target localization. Unlike existing deep learning approaches, MMWiLoc achieves computational efficiency and environmental adaptability without requiring extensive neural network training. The key contributions of our work include:
\begin{itemize}
	\item We provide a comprehensive multi-sensor dataset that synchronously captures human movement data across millimeter wave Wi-Fi, 2.4GHz Wi-Fi, and millimeter wave radar sensors. This dataset includes:
	\begin{itemize}
		\item Synchronized measurements from all three sensing modalities
		\item Ground truth location data for all movement patterns
		\item Diverse movement trajectories covering different scenarios
	\end{itemize}

	\item We develop MMWiLoc, an efficient localization system that achieves centimeter-level precision using only commercial millimeter wave Wi-Fi devices, demonstrating the potential of low-cost sensing solutions.
\end{itemize}


This paper is organized as follows: Section \ref{section2} introduces the sensing platform architecture and system design principles. Section \ref{section3} presents our enhanced expectation maximization algorithm and compressive sensing methodology. Section \ref{section4} validates the algorithm's effectiveness through extensive real-world experiments and detailed ablation studies. Finally, Section \ref{section5} summarizes our key findings and concludes the paper.

\section{System Design}
\label{section2}

\begin{figure}[ht]
	\centering
	\includegraphics[width = 0.45\textwidth]{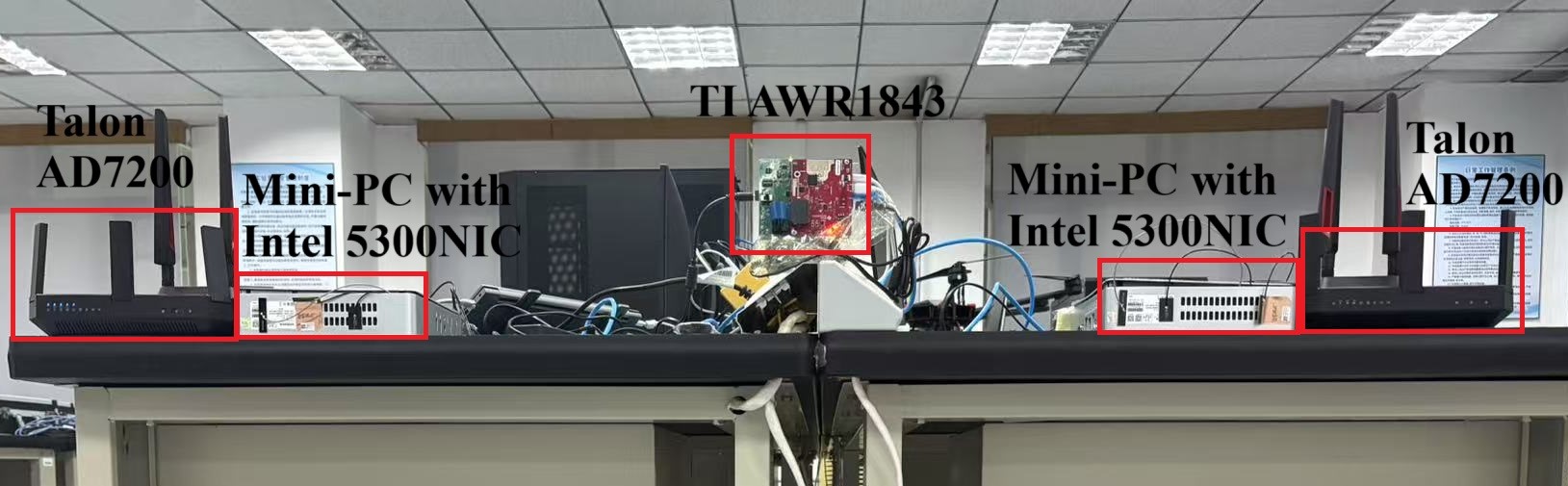}
	\caption{Sensor arrangement for multi-sensor platform. The system comprises two millimeter wave Wi-Fi devices (Talon AD7200 routers), two 2.4GHz Wi-Fi devices (Intel 5300 NICs), and a millimeter wave radar system (TI AWR1843 FMCW radar).}
	\label{systemdesign}
\end{figure}

Our work focuses on device-free Wi-Fi localization using millimeter wave Wi-Fi devices. In this section, we describe our multi-sensor data recording system that integrates radar, 2.4GHz Wi-Fi, and millimeter wave Wi-Fi, detailing the sensor specifications, experimental setup, and data processing methodologies.

\begin{figure}[ht]
	\centering
	\resizebox{0.45\textwidth}{!}{\begin{tikzpicture}[scale=1.5]
    \tikzset{
        mmwifi/.style={rectangle, draw, fill=blue!20, minimum size=0.3cm},
        wifi/.style={rectangle, draw, fill=green!20, minimum size=0.25cm},
        radar/.style={rectangle, draw, fill=red!20, minimum size=0.35cm},
        target/.style={circle, draw, fill=orange!20, minimum size=0.4cm},
        callout/.style={rectangle, draw, fill=white, rounded corners=2pt, font=\footnotesize}
    }
    
    \draw[->] (-4.5,0) -- (4.5,0) node[right] {x (m)};
    \draw[->] (0,0) -- (0,5.5) node[above] {y (m)};
    
    \draw[dashed, thick] (-4,0) rectangle (4,5);
    \path[pattern=north west lines, pattern color=black!10] (-4,0) rectangle (4,5);
    
    \node[mmwifi] (mmwifi1) at (-1.5,0) {};
    \node[mmwifi] (mmwifi2) at (1.5,0) {};
    \node[wifi] (wifi1) at (-0.5,0) {};
    \node[wifi] (wifi2) at (0.5,0) {};
    \node[radar] (radar) at (0,0) {};
    
    \begin{scope}[opacity=0.5]
        \draw[blue!30, fill=blue!10] (-1.5,0) -- ++(50:3.5) arc(50:120:3.5) -- cycle;
        \draw[blue!30, fill=blue!10] (1.5,0) -- ++(70:3.5) arc(70:140:3.5) -- cycle;
    \end{scope}
    
    \draw[->, blue, thick] (-1.5,0) -- ++(80:4);
    \draw[->, blue, thick] (1.5,0) -- ++(100:4);
    \node[blue] at (-1.3,2.5) {$\theta_1$};
    \node[blue] at (0.8,2.5) {$\theta_2$};
    
    \draw[dotted] plot[smooth] coordinates {
        (2,0.5) (-2,4) (-2,0.5) (2,4) (2,0.5)
    };
    \node[target] (target) at (-1,3.35) {};
    
    \draw[<->, green!60!black] (wifi1) -- (target);
    \draw[<->, green!60!black] (wifi2) -- (target);
    \draw[->, green!60!black] (target) -- ++(135:0.8) node[left] {$v$};
    
    \draw[->, red] (0,0) -- (target);
    \draw[red, dashed] (0,0) -- ++(90:1) arc(90:107:1);
    \node[red] at (-0.2,1.2) {$\theta_r$};
    \node[red] at (-0.6,1.5) {$r$};

    \node[above] at (-2.2,5.0) {\Large Measurement Area (5m $\times$ 3.5m)};
    
    \node[callout] at (-2.85,4.5) {\parbox{3cm}{\centering MMWi-Fi:\\ AoA from beamSNR\\ $N_{sector} \times 1$}};
    \node[callout] at (-0.15,4.5) {\parbox{4cm}{\centering 2.4GHz Wi-Fi:\\ ToF/DFS/AoA\\ $N_{subcarrier} \times N_{rx} \times N_{tx}$}};
    \node[callout] at (2.7,4.5) {\parbox{3.5cm}{\centering Radar:\\ Range/Angle\\ $N_{range} \times N_{doppler} \times N_{tx} \times N_{rx}$}};
    
    \begin{scope}[shift={(2.6,1.9)}]
        \node[mmwifi] at (0,0) {};
        \node[right] at (0.2,0) {MMWi-Fi};
        \node[wifi] at (0,-0.5) {};
        \node[right] at (0.2,-0.5) {Wi-Fi};
        \node[radar] at (0,-1) {};
        \node[right] at (0.2,-1) {Radar};
        \node[target] at (0,-1.5) {};
        \node[right] at (0.2,-1.5) {Target};
    \end{scope}
\end{tikzpicture}}
	\caption{Experimental setup showing sensor placement, measurement area, and sensing mechanisms. The figure illustrates how different sensors (MMWi-Fi, 2.4GHz Wi-Fi, and Radar) localize targets through AoA, ToF, DFS, and range measurements. Multiple beam sectors show the beam steering capability of MMWi-Fi devices.}
	\label{setup}
\end{figure}
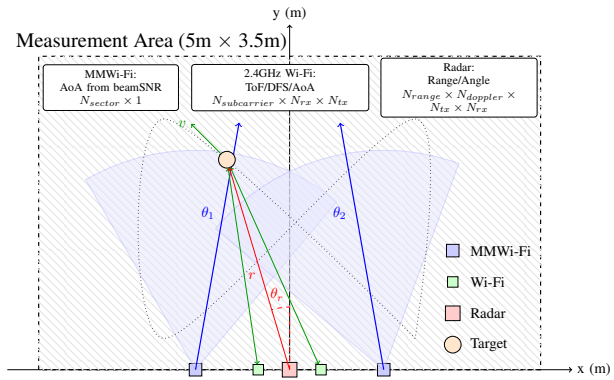

\begin{table*}[ht!]
	\caption{Comprehensive Comparison of Wi-Fi Sensing Datasets}
	\vspace{0.5em}\centering
	\begin{tabular}{p{2cm} c c c c c p{3cm} p{3cm}}
	\hline
	Dataset & \multicolumn{5}{c}{Sensors} & Scale & Focus Area \\
	\cline{2-6}
	& MMWi-Fi & 2.4GHz Wi-Fi & 5GHz Wi-Fi & RGB-D & mmWave Radar & & \\
	\hline
	MMWiLoc & \checkmark & \checkmark & -- & -- & \checkmark & 1-3 subjects, 6 patterns, 40 repetitions & Multi-sensor localization \\
	\hline
	MM-Fi\cite{yang_mm-fi_2023} & -- & -- & \checkmark & \checkmark & \checkmark & 40 subjects, 20+ actions, 4 scenarios & Multimodal sensing \\
	\hline
	Wi-Pose\cite{jiang_towards_2020} & -- & -- & \checkmark & \checkmark & -- & 16 subjects, 16 actions & Multimodal sensing \\
	\hline
	WiSense\cite{liao_wisense_2024} & -- & \checkmark & -- & -- & -- & 10 subjects, 6 activities & Activity recognition \\
	\hline
	WiARL-UIT\cite{thanh_public_2024} & -- & \checkmark & -- & -- & -- & 2 actions, 3 locations & Activity recognition \\
	\hline
	WiMANS\cite{huang_wimans_2024} & -- & \checkmark & \checkmark & -- & -- & 11,286 samples, 0-5 users & Multi-user activity sensing \\
	\hline
	NTU-Fi\cite{yang_sensefi_2023} & -- & \checkmark & -- & -- & -- & 6 activities, 14 gait patterns & High-resolution CSI, gait analysis \\
	\hline
	WiFi-80MHz\cite{meneghello_csi_2023} & -- & -- & \checkmark & -- & -- & 10 subjects, 3 applications & Wide-band sensing \\
	\hline
	Widar 3.0\cite{7znf-qp86-20} & -- & \checkmark & -- & -- & -- & 258K instances, 75 domains, 8,620 minutes & Hand gesture recognition \\
	\hline
	WiAR\cite{guo_huac_2018} & -- & \checkmark & -- & -- & -- & 16 activities, 10 volunteers, 30 repetitions & Activity/gesture recognition \\
	\hline
	UT-HAR\cite{yousefi_survey_2017} & -- & \checkmark & -- & -- & -- & 6 activities, continuous data & Continuous recognition \\
	\hline
	SignFi\cite{ma_signfi_2018} & -- & \checkmark & -- & -- & -- & 8,800 samples, 276 gestures & Sign language recognition \\
	\hline
	\end{tabular}
	\label{table_dataset}
\end{table*}

The experimental setup comprises two millimeter wave Wi-Fi devices (Talon AD7200 routers with QCA9500 802.11ad chipsets operating at 60GHz with 32-element phased arrays), two 2.4GHz Wi-Fi devices (Intel 5300 NICs with 3 antennas), and a millimeter wave radar system (TI AWR1843 FMCW radar operating at 77GHz with 3TX/4RX MIMO configuration), as shown in Figure \ref{systemdesign}. The sensors are arranged along the x-axis with the radar system positioned at the origin (0m, 0m). The millimeter wave Wi-Fi devices are located at (-0.75m, 0m) and (0.75m, 0m), and the 2.4GHz Wi-Fi devices are positioned at (-0.64m, 0m) and (0.64m, 0m). The target movement area spans 5m in length by 3.5m in width in front of the table as shown in Figure \ref{setup}, ensuring a clear line-of-sight throughout the experiments. The table is elevated 0.75m above the ground. 

\begin{figure}[htbp]
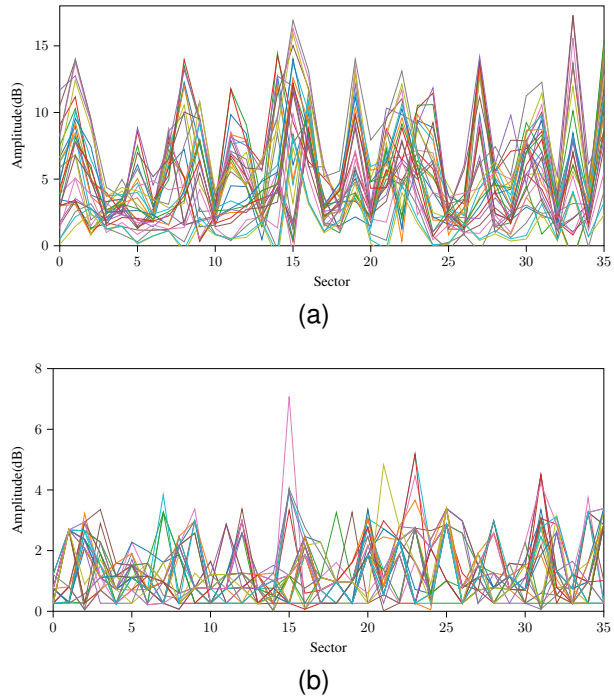

	\centering
	\subfloat[\label{fig:moving}]{
		\resizebox{0.45\textwidth}{!}{\input{figures/moving_sector.tex}}
	}
	\hfill
	\subfloat[\label{fig:static}]{
		\resizebox{0.45\textwidth}{!}{\input{figures/static_sector.tex}}
	}
	\caption{Recorded beamSNR data: (a) With moving subject (b) Without moving subject. Data is processed to highlight movement-related changes.}
	\label{exampledata}
\end{figure}

For the millimeter wave Wi-Fi system, we employ Talon AD7200 devices operating under the 802.11ad standard (60GHz with 2GHz bandwidth). These commercial mmWave Wi-Fi devices feature 36 sectors with distinct antenna array beam patterns. By using modified firmware provided in \cite{steinmetzer_compressive_2017}, we collect BeamSNR (Beam Signal-to-Noise Ratio) measurements from all sectors during target movement. Each measurement frame consists of 36 BeamSNR values, corresponding to specific beam patterns, as shown in Figure \ref{exampledata}. The raw BeamSNR data is preprocessed by subtracting the minimum value for each sector, thereby accentuating signal variations due to target presence. The devices operate in both transmitting and receiving modes, with measurements triggered via ping communications. Each measurement frame is structured as an $N_{sector} \times 1$ vector ($N_{sector}=36$), and the recording interval is optimized at 60ms to balance sampling density with system stability.

The 2.4GHz Wi-Fi system utilizes mini-PCs equipped with Intel IWL5300 wireless cards, operating at 2.4GHz with a bandwidth of 20MHz. CSI data are collected with a sampling rate of 20 million samples per second. The frame format is defined as $N_{subcarrier} \times N_{rx} \times N_{tx}$ (where $N_{subcarrier}=30$, $N_{rx}=3$, and $N_{tx}=1$) and the recording interval is 1ms.

The radar system is based on a Texas Instruments AWR1843 millimeter wave radar, transmitting frequency-modulated continuous wave (FMCW) signals within the 76–81GHz band and providing a 3.997GHz bandwidth. It samples at 4.76 million samples per second and outputs data in the format $N_{range} \times N_{doppler} \times N_{tx} \times N_{rx}$ (with $N_{range}=256$, $N_{doppler}=128$, $N_{rx}=4$, and $N_{tx}=2$), with a recording interval of 50ms.

System synchronization is achieved through NTP time servers and a central control computer. Our dataset comprises six movement patterns (hourglass, diamond, square, Z-shape, rectangle, and large hourglass), each recorded 40 times with 15-second intervals. Participants walked at 1.2 m/s along predefined paths marked with tape for ground-truth reference.

As shown in Table \ref{table_dataset}, MMWiLoc distinguishes itself from existing Wi-Fi sensing datasets through its unique multi-sensor approach. While most existing datasets rely on single-band Wi-Fi, MMWiLoc incorporates mmWave Wi-Fi along with traditional 2.4GHz Wi-Fi and mmWave radar, enabling comprehensive performance comparison across different sensing technologies.

For data processing, the radar system employs standard techniques including FFT for Range-Doppler mapping and CFAR detection, followed by angle estimation using the CAPON algorithm. The 2.4GHz Wi-Fi data is processed using the Widar2 algorithm \cite{qianWidar2PassiveHuman2018}. The processing methodology for the millimeter wave Wi-Fi data is detailed in following sections.

\section{Proposed Algorithm}
\label{section3}

In Section \ref{section2}, we introduced the multi-sensor platform and data collection process. This section focuses on the millimeter wave Wi-Fi sensing algorithm for device-free localization. The system estimates target Angle of Arrival (AoA) through beamSNR measurements and determines target position via geometric triangulation. The two MMWi-Fi devices alternate between transmitter and receiver roles, systematically measuring channel characteristics using different phased array beam patterns. Each device reports Signal-to-Noise Ratio (beamSNR) for all sectors. Given the known antenna radiation patterns, we estimate target AoA using compressive sensing algorithms.

Our approach begins with the calibration of the Measurement Matrix for millimeter wave Wi-Fi devices. The Measurement Matrix consists of angular patterns for different antenna array sectors. While the original Measurement Matrix for our Talon AD7200 devices has been previously estimated by \cite{steinmetzer_compressive_2017}, our experiments reveal inherent imperfections in these measurements. To address these limitations, we propose an enhanced expectation maximization algorithm in subsection \ref{MME} that effectively corrects the Measurement Matrix.

Building upon this corrected Measurement Matrix, we develop a Multi-scale compression sensing algorithm, detailed in subsection \ref{MSL}, that achieves two key objectives: First, it extracts accurate Angle of Arrival (AoA) information for moving targets, and second, it enables precise target localization through the fusion of data from two millimeter wave Wi-Fi devices.

\subsection{Measurement Matrix Calibration}
\label{MME}

We propose an advised Expectation-Maximization (EM) algorithm for millimeter wave Wi-Fi measurement matrix calibration. Given observed signal vectors $Y \in \mathbb{R}^{M \times N}$ and initial target vectors $X \in \mathbb{R}^{M \times K}$, where M is the number of frames, N is the number of output beamSNR sector, and K is the number of input radius with Angle of Arrival ground-truth which is equal to 192 in our cases, we estimate the measurement matrix $A \in \mathbb{R}^{N \times K}$ and static offset $p \in \mathbb{R}^K$.

The measurement model is:

\begin{equation}
    Y = (AH)^T + \epsilon, \quad \epsilon \sim \mathcal{N}(0, \sigma^2)
\end{equation}

The original value of $A$ is given by \cite{steinmetzer_compressive_2017}, which doesn't lay perfectly with real measurement matrix in our applications, thus the divergence in estimated AoA with ground-truth results as we will show in experiment section. Thus we propose this algorithm to correct and estimate the Measurement Matrix.

\begin{equation}
    H = X + p + \eta
\end{equation}
where $H$ represents hidden states and $\eta$ represents state uncertainty.

\subsubsection{Enhanced E-step}
The major enhancement appears in the E-step, where we estimate hidden states using a weighted regularization approach that improves stability:

\begin{equation}
    h_i = \operatorname{argmin}_h \{\|Y_i - (A^2h)^T\|^2 + \sigma\|h - (X_i + p)\|^2\}
\end{equation}

This optimization is solved through an iterative refinement process, providing stable convergence:

\begin{equation}
    h_i^{(t+1)} = \frac{h_i^{data} + \sigma(X_i + p)}{1 + \sigma}
\end{equation}

\subsubsection{M-step}
In the M-step, we incorporate several stability improvements. First, we implement square-root parameterization for guaranteed positive definiteness:

\begin{equation}
    A^2_{new} = Y^T H (H^T H)^{-1}
\end{equation}

\begin{equation}
    A = \sqrt{|A^2_{new}|}
\end{equation}

Second, we introduce static offset estimation:
\begin{equation}
    p = \frac{1}{M}\sum_{i=1}^M (h_i - X_i)
\end{equation}

Third, we employ noise variance estimation across multiple frames:
\begin{equation}
    \sigma = \frac{1}{MN}\sum_{i=1}^M \|Y_i - (A(h_i + p))^T\|^2
\end{equation}

We further achieve convergence monitoring through dual criteria. First, we track parameter changes:

\begin{equation}
    \|\Delta A\|_F^2 + \|\Delta p\|^2 < \epsilon
\end{equation}

Second, we monitor the log-likelihood:
\begin{equation}
    \mathcal{L} = -\frac{1}{2\sigma}\sum_{i=1}^M \|Y_i - Y_i^{pred}\|^2 - \frac{M}{2}\log(\sigma)
\end{equation}


\subsection{Multi-Scale Lasso for AoA Estimation}
\label{MSL}

With the calibrated measurement matrix $A$ from our enhanced EM algorithm, we now develop our multi-scale LASSO framework for precise AoA estimation, extracting target angular information $H$ from beamSNR measurements $Y$. Given the millimeter wave Wi-Fi beamSNR measurement model:
\begin{equation}
    Y = (AH)^T + \epsilon, \quad \epsilon \sim \mathcal{N}(0, \sigma^2)
\end{equation}
where $Y \in \mathbb{R}^{M \times N}$ represents the received signals (BeamSNR) from the receiver, where M is the number of frames, N is the number of output beamSNR sectors.  $A \in \mathbb{R}^{N \times K}$ is the measurement matrix, and $H$ represents the hidden states. The hidden states $H \in \mathbb{R}^{K \times 1}$ represent the target's Angle of Arrival (AoA) information, where K is the number of radius which is equal to 192 in our cases.

\begin{algorithm}
	\caption{Multi-Scale Adaptive Lasso for MMWiFi Signal Processing}
	\label{alg:mslasso}
	\begin{algorithmic}[1]
	\REQUIRE Measurement $Y$, measurement matrix $A$, scales $L$
	\ENSURE Reconstructed signal $\hat{H}$
	\STATE Initialize $\lambda_0$, $\beta$, weights $w_l$
	\FOR{$l = 1$ to $L$}
		\STATE Compute downsampled signals: $Y_l = D_l Y$
		\STATE Compute scale measurement matrix: $A_l = D_l A$
		\STATE Calculate scale-specific $\lambda_l = \lambda_0 \cdot 2^{l-1}$
		\STATE Solve Lasso optimization:
		\STATE $\hat{H}_l = \operatorname{argmin}_{H} \|Y_l - (A_lH)^T\|_2^2 + \lambda_l\|H\|_1$
	\ENDFOR
	\STATE Integrate scales: $\hat{H} = \sum_{l=1}^L w_l U_l\hat{H}_l$
	\RETURN $\hat{H}$
	\end{algorithmic}
\end{algorithm}

\subsubsection{Multi-Scale Decomposition}
We propose a multi-scale decomposition approach where the signal is analyzed at different resolution levels:
\begin{equation}
    Y_l = D_l Y, \quad l = 1,\ldots,L
\end{equation}
where $D_l$ is the down-sampling operator at level $l$, and $L$ is the total number of scales. The measurement matrices at each scale are:

\begin{equation}
    A_l = D_l A, \quad l = 1,\ldots,L
\end{equation}

At each scale $l$, we solve the following optimization problem:
\begin{equation}
    \hat{H}_l = \operatorname{argmin}_{H} \|Y_l - (A_l H)^T\|_2^2 + \lambda_l\|H\|_1
\end{equation}
where $\lambda_l$ is the scale-dependent regularization parameter:
\begin{equation}
    \lambda_l = \lambda_0 \cdot 2^{l-1}.
\end{equation}

This provides stronger regularization at coarser scales and finer detail recovery at higher resolutions.

The regularization parameter $\lambda_0$ is adaptively selected based on noise estimation:
\begin{equation}
    \lambda_0 = \alpha \cdot \hat{\sigma}\sqrt{\log(K)/M}
\end{equation}
where $\hat{\sigma}$ is the estimated noise level and $\alpha$ is a scaling factor.

The final solution combines information from all scales:
\begin{equation}
    \hat{H} = \sum_{l=1}^L w_l U_l\hat{H}_l
\end{equation}
where $U_l$ is the up-sampling operator and $w_l$ are scale-dependent weights:
\begin{equation}
    w_l = \frac{\exp(-\beta l)}{\sum_{k=1}^L \exp(-\beta k)}
\end{equation}

The complete algorithm is shown in Algorithm \ref{alg:mslasso}.

For practical implementation, we employ three scales: original resolution for detail preservation, 2x down-sampling for noise-detail balance, and 4x down-sampling for robust pattern extraction. This multi-scale approach effectively handles varying noise characteristics in millimeter wave Wi-Fi sensing. We estimate the AoA using peak detection on the reconstructed signal $\hat{H}$.

\subsection{AoA-Based Location Estimation}

Given AoA estimates $\theta_{0}$ and $\theta_{1}$ from two millimeter wave Wi-Fi devices at known positions $\mathbf{p}_{0} = (x_{0}, y_{0})$ and $\mathbf{p}_{1} = (x_{1}, y_{1})$, where these devices alternate between transmitter and receiver roles during the communication process, the target position is determined through geometric triangulation:

\begin{equation}
    \begin{cases}
    x = x_{0} + r_{0}\sin(\theta_{0}) \\
    y = y_{0} + r_{0}\cos(\theta_{0})
    \end{cases}
\end{equation}
where $r_{0}$ is obtained by solving the intersection of lines:
\begin{equation}
    \begin{bmatrix}
    \sin(\theta_{0}) & -\sin(\theta_{1}) \\
    \cos(\theta_{0}) & -\cos(\theta_{1})
    \end{bmatrix}
    \begin{bmatrix}
    r_{0} \\ r_{1}
    \end{bmatrix} =
    \begin{bmatrix}
    x_{1} - x_{0} \\
    y_{1} - y_{0}
    \end{bmatrix}
\end{equation}

\section{Experiment}
\label{section4}

This section presents comprehensive experimental results for MMWiLoc, our localization algorithm for millimeter wave Wi-Fi. We evaluate the system from two perspectives: real-world performance comparisons with commercial RF sensors (Subsection \ref{real}) and detailed algorithm analysis (Subsection \ref{ablation}). In the performance evaluation, we compare MMWiLoc against a standard radar system and 2.4GHz Wi-Fi devices with Widar2 algorithm. The MMWiLoc is validated using a comprehensive dataset collected from varying movement patterns. The algorithmic analysis assesses our improvements over baseline approaches—including ENET \cite{zou_regularization_2005}, OMP \cite{cai_orthogonal_2011}, and LASSO \cite{tibshirani_regression_1996}—and examines the effectiveness of the proposed Expectation Maximization algorithm.

\subsection{Real-world Comparison}
\label{real}

\begin{figure}[htb!]
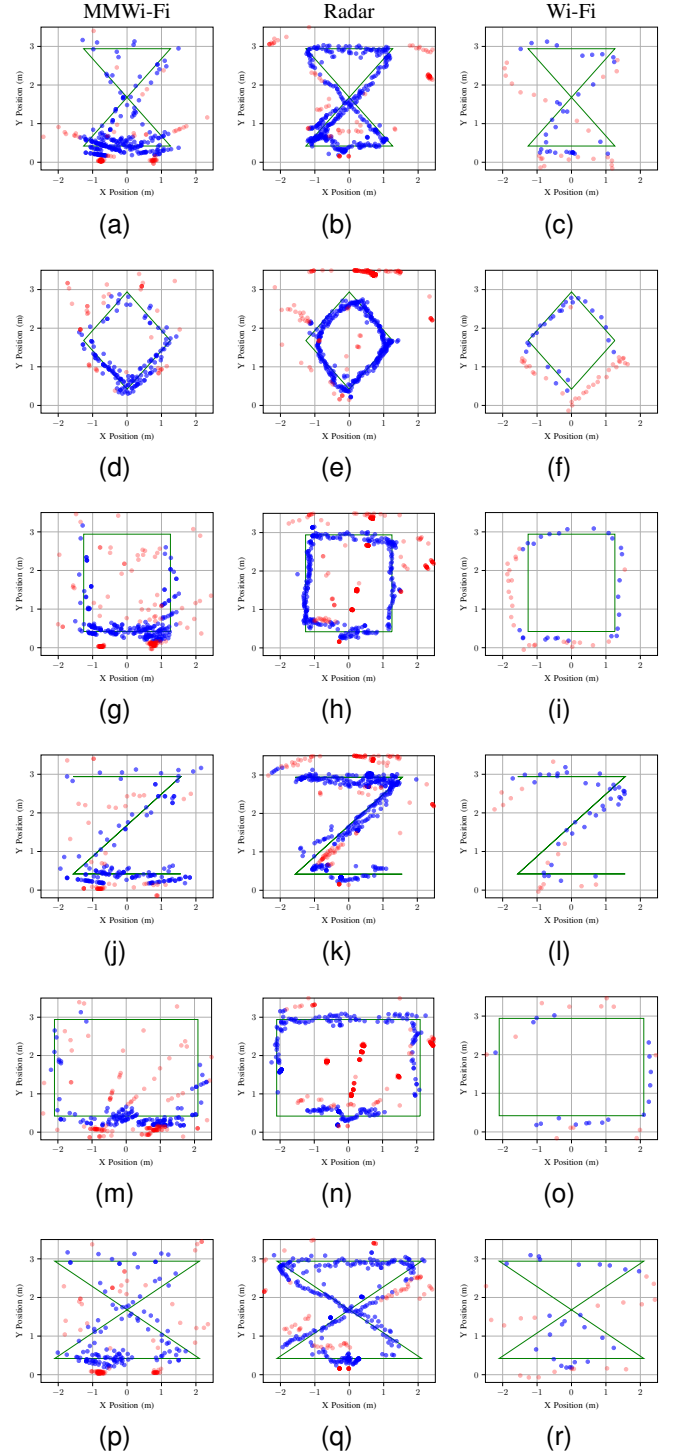

	\centering
	\subfloat[\label{fig:hourglass_mmwi}]{
		\resizebox{0.149\textwidth}{!}{\input{figures/hourglass_mmwifi.tex}}
	}
	\hfill
	\subfloat[\label{fig:hourglass_radar}]{
		\resizebox{0.149\textwidth}{!}{\input{figures/hourglass_radar.tex}}
	}
	\hfill
	\subfloat[\label{fig:hourglass_wi}]{
		\resizebox{0.149\textwidth}{!}{
\begin{tikzpicture}

    \definecolor{darkgray176}{RGB}{176,176,176}
    \definecolor{green01270}{RGB}{0,127,0}
    \definecolor{lightgray204}{RGB}{204,204,204}
    \definecolor{steelblue31119180}{RGB}{31,119,180}
    
    \begin{groupplot}[group style={group size=1 by 1}]
    \nextgroupplot[
    legend cell align={left},
    legend style={fill opacity=0.8, draw opacity=1, text opacity=1, draw=lightgray204},
    tick align=outside,
    title={\Huge Wi-Fi},
    tick pos=left,
    x grid style={darkgray176},
    xlabel={X Position (m)},
    xmajorgrids,
    xmin=-2.5, xmax=2.5,
    xtick style={color=black},
    y grid style={darkgray176},
    ylabel={Y Position (m)},
    ymajorgrids,
    ymin=-0.2, ymax=3.5,
    ytick style={color=black}
    ]
\addplot [draw=blue, fill=blue, mark=*, only marks, opacity=0.6]
table{%
x  y
0.03875966 0.26075906
0.034915976 0.25641096
0.019607777 0.24698994
0.053120382 0.24286824
0.09294814 0.22260173
-0.047214773 0.2507009
-0.14431284 0.2693881
-0.47148833 0.24676196
-0.7081799 0.27078673
-0.87885386 0.33546612
-0.7431431 0.7161107
-0.5634552 1.0986856
-0.33898053 1.5097148
0.19893736 1.7575482
0.63032126 2.0146215
1.2405679 2.5975966
1.2161117 2.7175372
0.8956133 2.7966335
0.59986126 2.7823913
0.3176564 2.7603488
-0.42603308 3.0267181
-0.7101663 3.1236885
-1.1608856 3.0863836
-1.4763863 2.9343696
-0.41672194 1.5535818
-0.1245265 1.3124268
0.83925 0.5308584
1.055862 0.22032528
-0.62518275 0.28358644
-0.91423714 0.25113955
-0.7383562 0.5179572
-0.46052262 0.9103023
-0.24248505 1.3187865
};

\addplot [draw=red, fill=red, mark=*, only marks, opacity=0.3]
table{%
x  y
-0.8866799 0.08657117
-0.9397284 -0.33116478
-0.94022256 0.16411388
0.91908777 2.2114015
1.3135833 2.356947
1.3342942 2.6444697
-1.8791173 2.5780306
-1.9422668 2.433443
-1.9114037 2.2431374
-1.7308031 2.0365384
-1.480417 1.8101077
-1.0903608 1.745266
-0.75771785 1.6043079
0.15687664 1.0723717
0.49444833 0.8203558
1.1676611 -0.07215553
1.1830881 -0.1581423
1.1744677 0.040628366
1.055404 0.13076179
0.63857335 0.08804416
0.18613228 0.13545893
-0.07865271 0.069727324
-0.29944044 0.1664327
-0.8474535 0.09851396
-0.95432943 -0.4265749
-0.9001346 -0.13665047
-0.7918553 -0.0988939
};
\addplot [thick, green01270]
table {%
0 0.42
-1.26 0.42
1.26 2.94
-1.26 2.94
1.26 0.42
0 0.42
};
     
\end{groupplot}
    
\end{tikzpicture}
    }
	}
	\hfill
	\subfloat[\label{fig:diamond_mmwi}]{
		\resizebox{0.149\textwidth}{!}{
\begin{tikzpicture}

\definecolor{darkgray176}{RGB}{176,176,176}
\definecolor{green01270}{RGB}{0,127,0}
\definecolor{lightgray204}{RGB}{204,204,204}
\definecolor{steelblue31119180}{RGB}{31,119,180}

\begin{groupplot}[group style={group size=1 by 1}]
\nextgroupplot[
legend cell align={left},
legend style={fill opacity=0.8, draw opacity=1, text opacity=1, draw=lightgray204},
tick align=outside,
tick pos=left,
x grid style={darkgray176},
xlabel={X Position (m)},
xmajorgrids,
xmin=-2.5, xmax=2.5,
xtick style={color=black},
y grid style={darkgray176},
ylabel={Y Position (m)},
ymajorgrids,
ymin=-0.2, ymax=3.5,
ytick style={color=black}
]
\addplot [draw=blue, fill=blue, mark=*, only marks, opacity=0.6]
table{%
x  y
1.16798147702414 1.56934657469598
0.849729741013608 1.27458075553243
1.07752716986482 1.45607777589775
0.808047243304347 1.11393113365783
0.996995505271629 1.24902019779121
0.704724258766738 1.15904799656287
0.867756730356778 1.2889437084261
0.852602409559721 1.31129445807177
0.698935393737263 0.979933783346109
0.64567004851964 1.14197663397733
0.673855139606772 0.833467763884683
0.834498654244309 1.29648136697288
0.809496424884491 0.912866683547681
0.504939668637082 1.02682702664076
0.566134959633349 0.770412704711784
0.703541410571884 1.18932862351981
0.716467979236905 0.858411600285792
0.531236329987898 0.866516728663556
0.294540761958572 0.706436480753703
0.404575933325455 0.780854645339644
0.287560525621989 0.607346370657444
0.381580378730816 0.662381831720984
0.0993014589958808 0.607210949628354
0.225956214574799 0.489020805400667
0.285984306019341 0.740680429243167
0.39163470347874 0.572037034730737
-0.143796828886221 0.433406947622171
0.0796730996469609 0.325045162839826
-0.012294609056632 0.369640805407387
0.141849879040185 0.446877762962172
-0.262706726428463 0.575227246855257
-0.0652784056969001 0.463086045042378
0.185252769028523 0.632523570229311
-0.26747286483458 0.727445626402398
-0.0993015038563586 0.607210949628354
-0.212338403551444 0.66926616501587
-0.362600213309134 0.817555201025233
-0.174507725667056 0.679342040802849
-0.435090464967793 0.894782064392124
-0.287299370087071 0.783194953631143
-0.844180414891003 1.20365835758237
-0.566806727536658 0.994232150963015
-0.694681104821333 1.06157318409698
-0.515274257584152 0.929742361755102
-0.694681104821333 1.06157318409698
-0.515274257584152 0.929742361755102
-0.919501010128363 1.29499765701819
-0.919501010128363 1.29499765701819
-1.2816758325748 1.57592923097576
-0.938844160663442 1.31000174175388
-1.2816758325748 1.57592923097576
-0.979118153026551 1.34124145076429
-1.38914981205624 1.89449085533369
-0.969529357827908 1.52286325414087
-1.31417634551094 1.82809225811426
-1.0083920055053 1.64087486324181
-1.07120749683337 2.03977391096515
-0.574570650313492 2.24204804891407
-0.574570650313492 2.24204804891407
-0.549161926585056 2.56678074847764
-0.261446294911463 2.65149279846598
-0.248283491433696 2.72293019872627
-0.0183930936873846 2.60976678791138
0.0392328242113882 2.8153287425271
0.0918540956287031 2.60689737015102
0.512744953808182 2.27028967912091
0.538378024012888 2.31637557944123
0.651566376369892 2.51987689555669
0.682942482839699 2.57628811355981
0.907168042893217 2.00866221871625
1.0515679554778 1.9150560776118
1.14923559309366 2.01887608805977
1.13558581709438 1.71418737323143
1.26070379756258 1.73455204151181
1.11114561703989 1.6055343043806
0.981585423089682 1.17109366275687
1.03956137017455 1.46426956008926
0.673855139606772 0.833467763884683
0.396758123134762 0.963506841718603
0.510807523685528 0.652424308980023
0.145531246432698 0.752425853026092
0.309538306792162 0.54827443648884
0.375196866411482 0.945391070562178
0.494657589665409 0.644067279680667
0.0492206276064901 0.653944770698865
0.21492462631089 0.499315128752462
0.164643523302157 0.748387146346891
0.317813763661673 0.552556667568236
0.184195405228019 0.667906160378437
0.292853347344874 0.539640484234295
-0.193142160688423 0.398127336603444
0.027970477820906 0.304789354771709
-0.0420526724935734 0.50614932126597
0.179902188106359 0.364312413260019
-0.291149742539336 0.439499145812743
-0.193142160688423 0.398127336603444
-0.0641301505777491 0.343667669634757
0.0390313114282 0.755755470119982
0.146815648051856 0.641181363966003
0.269454867788694 0.510816605774543
-0.240974922286583 0.344260256030819
-0.135839204085936 0.307736577730036
-0.0988153738456283 0.623722173783961
0.0310931931160045 0.528263424714219
0.141849879040185 0.446877762962172
-0.362600213309134 0.817555201025233
-0.193205915416066 0.693081749007284
0.0310931931160045 0.528263424714219
-0.314806066750052 0.782435350578932
-0.174507725667056 0.679342040802849
-0.330335410524634 0.710349620928337
-0.21692944284433 0.902307456087264
-0.50082208880443 1.10776299796018
-0.37121512149328 0.992979445646712
-0.527161206794838 0.990667726703967
-0.407502907849557 0.89785124091759
-0.537235708029828 0.94587996250256
-0.421596126361648 0.860906276261738
-0.825451202599093 1.1576649988368
-0.643027587337444 1.02361741043719
-0.694681104821333 1.06157318409698
-0.706877719468143 0.82752068067124
-0.680441039801379 1.33484157014525
-0.909621109467952 1.21951429951757
-0.643027587337444 1.02361741043719
-1.24443323787067 1.46553923611888
-0.92771668964267 1.23281120123523
-1.03445816862262 1.66519896828008
-1.03445816862262 1.66519896828008
-0.69336217212941 2.17597592769521
-0.297758594626734 2.45441741184955
-0.455007427907533 2.82277924288269
-0.220151839920172 2.87560720648208
0.66657748528123 2.13559614750145
0.66657748528123 2.13559614750145
0.971858223970882 1.40887296699159
1.43062481654389 1.78424876199812
1.22605227644034 1.61686177370701
};
\addplot [draw=red, fill=red, mark=*, only marks, opacity=0.3]
table{%
x  y
0.954808860985211 0.910512648189084
1.15292435253715 1.01632310635607
0.998758746498522 0.933985614679699
0.998758746498522 0.933985614679699
-0.0197053084979996 0.862079038294917
-1.10558827006323 1.05399214207824
-0.88377659013815 0.928000955652903
-1.10558827006323 1.05399214207824
-0.911185530059825 0.94356949947602
-1.35671016539421 1.9659141668686
-1.35671016539421 1.9659141668686
-1.72757257366633 3.1676141668978
-1.12812349314043 2.40120941250637
-1.59995248972061 2.63197454923406
-1.90855077475993 3.58758449306936
-1.15476423051405 2.57038634778848
-1.90855077475993 3.58758449306936
-1.15476423051405 2.57038634778848
-0.524558167076596 2.88122560509095
-0.819405477067539 5.33032158209821
0.349407516987373 3.40444061370061
-2.25530915631104 -4.66136120784565
0.427434124235817 3.08662762010108
0.427434124235817 3.08662762010108
0.460168681764937 3.17244078544019
0.610664198530404 3.5669627297934
1.53353582462797 5.98625850896059
1.21161905720838 3.52679599611687
0.654186754010702 3.68105675868872
0.704123046074854 1.76255036107504
0.914977597880215 0.834267618387846
1.52964849713192 1.96655967176343
0.914977597880215 0.834267618387846
0.902306058151676 0.967193751930099
-0.350762829248162 0.382400083901005
-1.09244283932093 1.01502800759305
-0.879402968268769 0.897661307667827
-1.35671016539421 1.9659141668686
-1.35671016539421 1.9659141668686
-1.68567832289597 3.0318644743613
-1.15509724112247 2.3714119353394
-1.72757257366633 3.1676141668978
-1.16914709656245 2.45365871258257
-0.676634575202582 2.81863106346414
-0.680934627126051 2.65342754073727
-0.586814458732291 3.13154260986393
0.396382207369348 3.00522557464216
0.427434124235817 3.08662762010108
0.427434124235817 3.08662762010108
0.877886966124261 2.45416087044724
1.38713257904548 3.22188640244058
};
    
\addplot [thick, green01270]
table {%
1.26 1.68
0 0.42
-1.26 1.68
0 2.94
1.26 1.68
};
    
\end{groupplot}

\end{tikzpicture}}
	}
	\hfill
	\subfloat[\label{fig:daimond_radar}]{
		\resizebox{0.149\textwidth}{!}{\input{figures/diamond_radar.tex}}
	}
	\hfill
	\subfloat[\label{fig:diamond_wi}]{
		\resizebox{0.149\textwidth}{!}{
\begin{tikzpicture}

    \definecolor{darkgray176}{RGB}{176,176,176}
    \definecolor{green01270}{RGB}{0,127,0}
    \definecolor{lightgray204}{RGB}{204,204,204}
    \definecolor{steelblue31119180}{RGB}{31,119,180}
    
    \begin{groupplot}[group style={group size=1 by 1}]
    \nextgroupplot[
    legend cell align={left},
    legend style={fill opacity=0.8, draw opacity=1, text opacity=1, draw=lightgray204},
    tick align=outside,
    tick pos=left,
    x grid style={darkgray176},
    xlabel={X Position (m)},
    xmajorgrids,
    xmin=-2.5, xmax=2.5,
    xtick style={color=black},
    y grid style={darkgray176},
    ylabel={Y Position (m)},
    ymajorgrids,
    ymin=-0.2, ymax=3.5,
    ytick style={color=black}
    ]
    \addplot [draw=blue, fill=blue, mark=*, only marks, opacity=0.6]
    table{%
    x  y
    -0.1224618 0.383796
    -0.1029054 0.6599035
    -0.36553758 0.87932944
    -0.6883779 0.9954939
    -1.2754455 1.6193525
    -1.1222949 1.815947
    -0.90403223 2.0665138
    -0.60853606 2.252656
    -0.3441534 2.4501336
    -0.24038549 2.5919251
    -0.16452627 2.678128
    -0.00702114 2.785381
    0.40072662 2.6926866
    0.86829656 2.2827034
    0.9985643 2.021392
    1.0761466 1.6860198
    1.2286642 1.3828869
    -0.45114073 0.57306004
    -1.3162528 1.3917581
    -1.2307836 1.5653304
    -1.0128314 1.8800287
    -0.7844201 2.1050487
    -0.5325215 2.305201
    -0.28453925 2.456529
    0.035885517 2.669481
    0.19808799 2.776694
    0.6057124 2.4565966
    };
    \addplot [draw=red, fill=red, mark=*, only marks, opacity=0.3]
    table{%
    x  y
    1.3785698 1.2438494
    1.5556165 1.0968663
    1.6308671 1.0665802
    1.5119636 1.2263398
    1.4889202 1.2125868
    1.4067479 1.1290059
    1.2041306 1.0611409
    1.0184435 0.7353069
    0.7612165 0.607951
    0.49684015 0.4785728
    0.34816268 0.32051343
    0.11300696 0.17624885
    -0.08725336 -0.13525209
    -1.0785623 0.8243606
    -1.2173026 1.015137
    -1.3644525 1.1599008
    -1.431529 1.2178339
    1.3827344 1.152097
    1.3030401 0.99156153
    1.0848684 0.9182559
    0.7912132 0.732006
    0.6341797 0.53444964
    0.42837232 0.3506376
    0.21826468 0.18282753
    0.0050226413 0.09159174
    0.018647918 0.00184099
    -0.2596771 0.3112884
    -0.7407559 0.70368916
    -1.1203284 0.8574821
    -1.3078592 1.0057443
    -1.3902385 1.1118461
    -1.4406744 1.1398116
    -0.021983748 2.5437353
    };
     
    \addplot [thick, green01270]
    table {%
    1.26 1.68
    0 0.42
    -1.26 1.68
    0 2.94
    1.26 1.68
    };
     
    \end{groupplot}
    
\end{tikzpicture}
    }
	}
	\hfill
	\subfloat[\label{fig:square_mmwi}]{
		\resizebox{0.149\textwidth}{!}{\input{figures/square_mmwifi.tex}}
	}
	\hfill
	\subfloat[\label{fig:square_radar}]{
		\resizebox{0.149\textwidth}{!}{\input{figures/square_radar.tex}}
	}
	\hfill
	\subfloat[\label{fig:square_wi}]{
		\resizebox{0.149\textwidth}{!}{
\begin{tikzpicture}

    \definecolor{darkgray176}{RGB}{176,176,176}
    \definecolor{green01270}{RGB}{0,127,0}
    \definecolor{lightgray204}{RGB}{204,204,204}
    \definecolor{steelblue31119180}{RGB}{31,119,180}
    
    \begin{groupplot}[group style={group size=1 by 1}]
    \nextgroupplot[
    legend cell align={left},
    legend style={
      fill opacity=0.8,
      draw opacity=1,
      text opacity=1,
      at={(0.03,0.97)},
      anchor=north west,
      draw=lightgray204
    },
    tick align=outside,
    tick pos=left,
    x grid style={darkgray176},
    xlabel={X Position (m)},
    xmajorgrids,
    xmin=-2.5, xmax=2.5,
    xtick style={color=black},
    y grid style={darkgray176},
    ylabel={Y Position (m)},
    ymajorgrids,
    ymin=-0.2, ymax=3.5,
    ytick style={color=black}
    ]
    \addplot [draw=blue, fill=blue, mark=*, only marks, opacity=0.6]
    table{%
    x  y
    -0.46130824 0.24060088
    -0.7571157 0.21366087
    -1.4051632 0.26312244
    -1.5071179 0.4541612
    -1.4100522 2.5527873
    -1.098123 2.6925027
    -0.6375577 2.9431136
    -0.360227 2.9623005
    -0.1188338 3.0673475
    0.64704424 3.0872617
    1.0778586 2.9684176
    1.1885456 2.8307974
    1.438329 2.5850072
    1.4665877 2.4228277
    1.3969277 1.7450885
    1.3557526 1.3200817
    1.381935 0.8554553
    1.3403264 0.49382102
    1.3133745 0.30619138
    1.1602252 0.29853237
    -0.08676937 0.18201159
    -0.40222585 0.25432655
    -0.69738334 0.2594706
    -1.4119931 0.24947898
    -1.2689314 2.7068038
    -0.8881686 2.8755984
    };
    \addplot [draw=red, fill=red, mark=*, only marks, opacity=0.3]
    table{%
    x  y
    1.082235 0.1586365
    0.6234226 0.063687086
    0.24460019 0.13068001
    -0.22405525 0.1522621
    -0.13443032 0.13260992
    -1.0584013 0.07541397
    -1.2280074 0.06554287
    -1.7513006 0.79344684
    -1.7570331 1.2979954
    -1.7356437 1.7438521
    -1.8642359 1.9035014
    -1.8570768 2.176074
    -1.6228037 2.3540597
    1.5478067 2.0229356
    0.078311734 0.13992453
    0.021549214 0.025651984
    -1.08193 0.060040846
    -1.7468808 -0.33788896
    -1.3870686 -0.052414175
    -1.6782289 0.581442
    -1.7190539 1.0855592
    -1.8232825 1.4479703
    -1.8304479 1.7324128
    -1.8050416 2.0055037
    -1.6127684 2.296759
    -1.5111034 2.5634744
    };
     
    \addplot [thick, green01270]
    table {%
    1.26 0.42
    0 0.42
    -1.26 0.42
    -1.26 2.94
    1.26 2.94
    1.26 0.42
    };
     
    \end{groupplot}
    
\end{tikzpicture}
    }
	}
	\hfill
	\subfloat[\label{fig:zshape_mmwi}]{
		\resizebox{0.149\textwidth}{!}{\input{figures/zshape_mmwifi.tex}}
	}
	\hfill
	\subfloat[\label{fig:zshape_radar}]{
		\resizebox{0.149\textwidth}{!}{\input{figures/zshape_radar.tex}}
	}
	\hfill
	\subfloat[\label{fig:zshape_wi}]{
		\resizebox{0.149\textwidth}{!}{
\begin{tikzpicture}

    \definecolor{darkgray176}{RGB}{176,176,176}
    \definecolor{green01270}{RGB}{0,127,0}
    \definecolor{lightgray204}{RGB}{204,204,204}
    \definecolor{steelblue31119180}{RGB}{31,119,180}
    
    \begin{groupplot}[group style={group size=1 by 1}]
    \nextgroupplot[
    legend cell align={left},
    legend style={
      fill opacity=0.8,
      draw opacity=1,
      text opacity=1,
      at={(0.03,0.97)},
      anchor=north west,
      draw=lightgray204
    },
    tick align=outside,
    tick pos=left,
    x grid style={darkgray176},
    xlabel={X Position (m)},
    xmajorgrids,
    xmin=-2.5, xmax=2.5,
    xtick style={color=black},
    y grid style={darkgray176},
    ylabel={Y Position (m)},
    ymajorgrids,
    ymin=-0.2, ymax=3.5,
    ytick style={color=black}
    ]
    \addplot [draw=blue, fill=blue, mark=*, only marks, opacity=0.6]
    table{%
    x  y
    0.698979 0.35705715
    -0.3056279 0.24208614
    -0.66338044 0.36815304
    -0.8035939 0.44964087
    -0.38164946 1.1943879
    -0.1420044 1.5517727
    0.10571974 1.7590164
    0.5309201 1.9919996
    0.94205177 2.085173
    1.365328 2.2183616
    1.393785 2.4073327
    1.5132419 2.567987
    1.539029 2.5577624
    1.2636061 2.6286454
    1.1197696 2.6179733
    0.7734796 2.7319367
    0.4098353 2.7684276
    -0.40391326 3.0191195
    -0.41247183 2.9534545
    -0.91058755 2.9925945
    -1.3548155 2.9871395
    -1.7078674 2.8921955
    -1.0732331 2.7920544
    -0.6543029 2.9905453
    -0.13359964 3.0476735
    -0.17695561 3.1829302
    0.82301253 3.017239
    1.2211611 2.7276578
    1.3834779 2.659989
    1.5417154 2.4910913
    1.4365876 2.3713846
    1.1390111 2.3657448
    0.90673083 2.1823797
    0.7087094 1.9355955
    0.37609035 1.6335697
    0.020063527 1.4649975
    -0.58225405 0.6510323
    -0.78577185 0.36588687
    };
    \addplot [draw=red, fill=red, mark=*, only marks, opacity=0.3]
    table{%
    x  y
    0.7112486 0.14329109
    0.29202396 0.10183585
    -0.5321492 3.326317
    -0.38121626 3.5027266
    -0.91715664 0.06856927
    -0.97050637 -0.2868763
    -0.96305496 -0.037752014
    -0.81099313 0.15108773
    -0.6897979 0.7078583
    -2.0309265 2.5295584
    -2.23559 2.0905995
    -1.98786 2.2043624
    -1.7618798 2.379329
    -1.4384298 2.6209311
    -0.19528428 1.2002233
    -0.38828284 0.9322741
    -0.8751046 0.05353193
    -0.8730669 -0.2255757
    };
     
    \addplot [thick, green01270]
    table {%
    1.56 0.42
    0 0.42
    -1.56 0.42
    1.56 2.94
    -1.56 2.94
    1.56 2.94
    -1.56 0.42
    0 0.42
    1.56 0.42
    };
     
    \end{groupplot}
    
\end{tikzpicture}
    }
	}
	\hfill
	\subfloat[\label{fig:rectangle_mmwi}]{
		\resizebox{0.149\textwidth}{!}{\input{figures/rectangle_mmwifi.tex}}
	}
	\hfill
	\subfloat[\label{fig:rectangle_radar}]{
		\resizebox{0.149\textwidth}{!}{\input{figures/rectangle_radar.tex}}
	}
	\hfill
	\subfloat[\label{fig:rectangle_wi}]{
		\resizebox{0.149\textwidth}{!}{
\begin{tikzpicture}

    \definecolor{darkgray176}{RGB}{176,176,176}
    \definecolor{green01270}{RGB}{0,127,0}
    \definecolor{lightgray204}{RGB}{204,204,204}
    \definecolor{steelblue31119180}{RGB}{31,119,180}
    
    \begin{groupplot}[group style={group size=1 by 1}]
    \nextgroupplot[
    legend cell align={left},
    legend style={fill opacity=0.8, draw opacity=1, text opacity=1, draw=lightgray204},
    tick align=outside,
    tick pos=left,
    x grid style={darkgray176},
    xlabel={X Position (m)},
    xmajorgrids,
    xmin=-2.5, xmax=2.5,
    xtick style={color=black},
    y grid style={darkgray176},
    ylabel={Y Position (m)},
    ymajorgrids,
    ymin=-0.2, ymax=3.5,
    ytick style={color=black}
    ]
    \addplot [draw=blue, fill=blue, mark=*, only marks, opacity=0.6]
    table{%
    x  y
    1.0688683 0.31705263
    0.5111256 0.23334703
    0.12653358 0.18979174
    -1.0183861 0.17895053
    -2.2071683 2.0538776
    -1.1069021 2.8409932
    -1.0125204 2.9498825
    -0.49671972 3.018108
    2.2447171 2.3158698
    2.2654333 1.9534165
    2.3058484 1.5531722
    2.2861173 1.206447
    2.2468772 0.7807561
    2.101743 0.44422585
    0.8940426 0.24016434
    0.43312532 0.22311082
    -0.3664789 0.35529244
    -0.8895002 0.21278858
    };
    \addplot [draw=red, fill=red, mark=*, only marks, opacity=0.3]
    table{%
    x  y
    -0.026720535 0.10864671
    -0.86125433 3.3357437
    -2.7072887 0.77200085
    -2.878338 1.0521874
    -2.8631155 1.3551198
    -2.7286828 1.5841688
    -2.677455 1.7013885
    -2.465115 2.0042615
    -1.6237437 2.4656067
    -0.061611034 3.2439256
    0.86458826 3.2516084
    1.2279236 3.2427032
    1.0209525 3.4621284
    0.42907292 3.6698604
    2.4491851 1.988128
    1.9893214 0.052026242
    1.8934686 -0.15977488
    0.051874295 0.11061166
    -0.42965782 -0.16328746
    -1.3901521 -0.40846878
    };
     
    \addplot [thick, green01270]
    table {%
    2.1 0.42
    0 0.42
    -2.1 0.42
    -2.1 2.94
    2.1 2.94
    2.1 0.42
    };
     
    \end{groupplot}
    
\end{tikzpicture}
    }
	}
	\hfill
	\subfloat[\label{fig:lhour_mmwi}]{
		\resizebox{0.149\textwidth}{!}{\input{figures/lhour_mmwifi.tex}}
	}
	\hfill
	\subfloat[\label{fig:lhour_radar}]{
		\resizebox{0.149\textwidth}{!}{\input{figures/lhour_radar.tex}}
	}
	\hfill
	\subfloat[\label{fig:lhour_wi}]{
		\resizebox{0.149\textwidth}{!}{
\begin{tikzpicture}

\definecolor{darkgray176}{RGB}{176,176,176}
\definecolor{green01270}{RGB}{0,127,0}
\definecolor{lightgray204}{RGB}{204,204,204}
\definecolor{steelblue31119180}{RGB}{31,119,180}

\begin{groupplot}[group style={group size=1 by 1}]
\nextgroupplot[
legend cell align={left},
legend style={fill opacity=0.8, draw opacity=1, text opacity=1, draw=lightgray204},
tick align=outside,
tick pos=left,
x grid style={darkgray176},
xlabel={X Position (m)},
xmajorgrids,
xmin=-2.5, xmax=2.5,
xtick style={color=black},
y grid style={darkgray176},
ylabel={Y Position (m)},
ymajorgrids,
ymin=-0.2, ymax=3.5,
ytick style={color=black}
]
\addplot [draw=blue, fill=blue, mark=*, only marks, opacity=0.6]
table{%
x  y
0.0045724353 0.1864495
-0.04718735 0.19235471
-0.1475305 0.18535718
-1.0285745 0.6549669
-0.6256219 1.0402097
-0.2740846 1.3048604
0.25739983 1.545869
1.9628245 2.6790316
1.5644604 2.7253504
1.1121596 2.8290787
0.82419366 2.854458
0.2132779 2.850132
-0.91530055 2.9845622
-0.94672066 3.062092
-1.1897672 3.0958252
-1.8851018 2.8757808
-0.63686275 1.5270578
-0.20045477 1.3852684
0.37820742 1.2297318
0.898091 0.955693
1.228125 0.69434565
0.47249573 0.33729386
0.09520881 0.32465136
-0.29460186 0.22998178
-0.51442695 0.40365243
};
\addplot [draw=red, fill=red, mark=*, only marks, opacity=0.3]
table{%
x  y
0.07018157 0.14902014
0.006760303 0.14964646
-0.49253994 0.07538243
-0.954634 -0.06803353
-1.4877591 -0.020987678
-2.1322286 -0.2975238
-1.5392128 -0.24478091
-1.2395113 -0.07027317
0.7635131 1.7872336
1.5065085 1.8697553
2.0426023 1.822917
2.4218524 1.9423231
2.607949 2.0175443
2.3923361 2.3531787
2.1783223 2.5429149
-2.4833772 2.2940674
-2.8441844 1.8286979
-2.8390863 1.733331
-2.9175713 1.3357795
-2.619741 1.4672982
-2.3329866 1.4180498
-1.9573979 1.2769321
-1.4705372 1.4599006
-1.0696604 1.4761642
1.5852407 0.1408811
1.8545485 -0.5608535
1.033037 0.16497223
};
    
\addplot [thick, green01270]
table {%
0 0.42
-2.1 0.42
2.1 2.94
-2.1 2.94
2.1 0.42
0 0.42
};
    
\end{groupplot}

\end{tikzpicture}}
	}
	\caption{The localization results of three different sensors. The green line represents the ground truth of the moving trajectory. The blue dots represent the localization point within 0.25m of the ground truth. The red dots represent a localization point over 0.25m of the ground truth. (a) (d) (g) (j) (m) (p) are millimeter wave Wi-Fi localization results. (b) (e) (h) (k) (n) (q) are radar localization results. (c) (f) (i) (l) (o) (r) are 2.4GHz Wi-Fi localization results.}
	\label{fig:figure_2}
\end{figure}

We evaluate system performance across multiple movement patterns to thoroughly assess localization capabilities under different scenarios, using same expectation maximization and multi-scale LASSO parameters. For expectation maximization estimation, we initialize with noise variance $\sigma(0)=1$ and convergence threshold $\epsilon=10^{-4}$. In the multi-scale LASSO framework, we employ a three-scale decomposition ($L=3$) with scale weight factor $\beta=0.75$ and adaptive noise scaling $\alpha=1.0$. The Widar2 uses a sliding window size of 20. The radar uses the CFAR algorithm and peak detection to detect the moving target.

Figure \ref{fig:figure_2} presents results for four distinct trajectory patterns. The hourglass pattern tests the system's ability to localize rapidly changing trajectories at varying distances from the sensors. The diamond pattern evaluates performance in scenarios with sharp turning points and straight-line segments. The square pattern examines the system's performance with vertical movements and parallel paths relative to the sensors. Lastly, the Z-shape pattern combines diagonal movements with sharp turns to assess overall localization stability.

In all cases, green lines indicate ground-truth trajectories, while blue and red dots represent localization points that fall within or exceed the 0.25m error threshold, respectively. Notably, MMWiLoc demonstrates strong performance when targets are within 3 meters of the sensors, achieving accuracy comparable to dedicated radar systems despite relying solely on angle information. For instance, in the diamond pattern, the radar system exhibits highly stable performance with most points (blue dots) closely following the ground-truth trajectory across the entire movement path. In contrast, while MMWiLoc maintains reasonable accuracy, it shows slightly more variation in the distribution of localization points, particularly at the trajectory's corners. Meanwhile, the 2.4GHz Wi-Fi implementation exhibits considerably less stability, with numerous points (red dots) deviating significantly from the ground-truth—especially in lower portions of the pattern and at turning points. These deviations likely arise from the challenges of extracting reliable ToF and DFS information from lower bandwidth signals. 

It is important to note that our current implementation focuses on single-target scenarios, as we are presently able to obtain reliable AoA information from millimeter wave Wi-Fi only. We plan to address multi-target tracking capabilities in future work.


\begin{figure}[htp]
	\centering
	\subfloat[\label{fig:hourglass_error}]{
		\resizebox{0.21\textwidth}{!}{
\begin{tikzpicture}

\definecolor{crimson2143940}{RGB}{214,39,40}
\definecolor{darkgray176}{RGB}{176,176,176}
\definecolor{darkturquoise23190207}{RGB}{23,190,207}
\definecolor{lightgray204}{RGB}{204,204,204}
\definecolor{orchid227119194}{RGB}{227,119,194}
\definecolor{steelblue31119180}{RGB}{31,119,180}

\begin{groupplot}[group style={group size=1 by 1}]
\nextgroupplot[
legend cell align={left},
legend style={
    fill opacity=0.8,
    draw opacity=1,
    text opacity=1,
    at={(0.98,0.40)},  
    anchor=north east,
    draw=black!20
},
tick align=outside,
tick pos=left,
x grid style={darkgray176},
xlabel={ Mean Distance (m)},
xmajorgrids,
xmin=0.05, xmax=0.25,
xtick={0.05, 0.10, 0.15, 0.20, 0.25},
xticklabels={0.05, 0.10, 0.15, 0.20, 0.25},
xtick style={color=black},
y grid style={darkgray176},
ylabel={Cumulative Probability},
ymajorgrids,
ymin=0, ymax=1.00,
ytick style={color=black},
ytick={0,0.2,0.4,0.6,0.8,1},
yticklabels={0,0.2,0.4,0.6,0.8,1}
]
\addplot [thick, crimson2143940, mark=*, mark size=1.5, mark options={solid}]
table{%
0.0953018181238052 0.0263157894736842
0.0978662601328852 0.0526315789473684
0.101601785241959 0.0789473684210526
0.102269044295193 0.105263157894737
0.10280975990777 0.131578947368421
0.104453358473697 0.157894736842105
0.104509653568816 0.184210526315789
0.105180832001938 0.210526315789474
0.105253375473239 0.236842105263158
0.105992332222538 0.263157894736842
0.107557831686 0.289473684210526
0.108188579280199 0.315789473684211
0.110943679083007 0.342105263157895
0.111960859043462 0.368421052631579
0.113114116965693 0.394736842105263
0.114898442665959 0.421052631578947
0.115443682529317 0.447368421052632
0.115630220327291 0.473684210526316
0.11978180161915 0.5
0.120773980437058 0.526315789473684
0.123267473613735 0.552631578947368
0.12732119613393 0.578947368421053
0.127602661050185 0.605263157894737
0.129895597705807 0.631578947368421
0.132497283897668 0.657894736842105
0.135105418143572 0.684210526315789
0.135453765725692 0.710526315789474
0.136073277357164 0.736842105263158
0.137994888406574 0.763157894736842
0.141139351306807 0.789473684210526
0.141504977458713 0.815789473684211
0.143926937866648 0.842105263157895
0.145128485626995 0.868421052631579
0.164525710079087 0.894736842105263
0.177292343815833 0.921052631578947
0.182436745197116 0.947368421052632
0.190177175785 0.973684210526316
0.245749799653155 1
};
\addlegendentry{MMWi-Fi}
\addplot [thick, steelblue31119180, mark=x, mark size=1.5, mark options={solid}]
table {%
0.0803901561558329 0.0263157894736842
0.0817389655197798 0.0526315789473684
0.0836349467060748 0.0789473684210526
0.0845779321897215 0.105263157894737
0.0899518348290108 0.131578947368421
0.0899846897689347 0.157894736842105
0.0915858412452654 0.184210526315789
0.0949306942138688 0.210526315789474
0.0957653656073998 0.236842105263158
0.0958916751332121 0.263157894736842
0.0966871695978031 0.289473684210526
0.0969654048130988 0.315789473684211
0.0971260315895923 0.342105263157895
0.0983466174576689 0.368421052631579
0.0986572422876437 0.394736842105263
0.0987286210325773 0.421052631578947
0.0993930866665856 0.447368421052632
0.103368843032787 0.473684210526316
0.104703820399254 0.5
0.105652636206225 0.526315789473684
0.105988460082886 0.552631578947368
0.107216483968864 0.578947368421053
0.107552776060791 0.605263157894737
0.107625631962608 0.631578947368421
0.108787040886059 0.657894736842105
0.110786472970772 0.684210526315789
0.111515743449672 0.710526315789474
0.111909842797601 0.736842105263158
0.11243079154487 0.763157894736842
0.113412746480393 0.789473684210526
0.114327262952084 0.815789473684211
0.115825432718702 0.842105263157895
0.119624407565886 0.868421052631579
0.12207449091646 0.894736842105263
0.123631723428996 0.921052631578947
0.12458008870277 0.947368421052632
0.133398344310169 0.973684210526316
0.135273470745157 1
};
\addlegendentry{Radar}
\addplot [thick, orchid227119194, mark=diamond, mark size=1.5, mark options={solid}]
table {%
0.132397753000663 0.0263157894736842
0.13398209416884 0.0526315789473684
0.136738976748341 0.0789473684210526
0.137720579187351 0.105263157894737
0.137752769998412 0.131578947368421
0.140134470676489 0.157894736842105
0.143247472265574 0.184210526315789
0.143809807335655 0.210526315789474
0.143885613359497 0.236842105263158
0.146195257180362 0.263157894736842
0.146631234574407 0.289473684210526
0.148692267780015 0.315789473684211
0.149055591227787 0.342105263157895
0.149426428593875 0.368421052631579
0.151258030713381 0.394736842105263
0.151913430587724 0.421052631578947
0.15308838024887 0.447368421052632
0.1545182823185 0.473684210526316
0.154947826342617 0.5
0.156510035680589 0.526315789473684
0.157756539992973 0.552631578947368
0.158339867313936 0.578947368421053
0.162324915004298 0.605263157894737
0.16264694281081 0.631578947368421
0.163183679964701 0.657894736842105
0.164416139530626 0.684210526315789
0.165073975368896 0.710526315789474
0.173113515 0.736842105263158
0.180529363333333 0.763157894736842
0.197977245365483 0.789473684210526
0.205158642334308 0.815789473684211
0.214631895 0.842105263157895
0.217302822037817 0.868421052631579
0.218109447727273 0.894736842105263
0.221079882619048 0.921052631578947
0.234695549090909 0.947368421052632
0.236473501481481 0.973684210526316
0.236535054464286 1
};
\addlegendentry{Wi-Fi}
\end{groupplot}
\end{tikzpicture}}
	}
	\subfloat[\label{fig:diamond_error}]{
		\resizebox{0.21\textwidth}{!}{
\begin{tikzpicture}

\definecolor{crimson2143940}{RGB}{214,39,40}
\definecolor{darkgray176}{RGB}{176,176,176}
\definecolor{darkturquoise23190207}{RGB}{23,190,207}
\definecolor{lightgray204}{RGB}{204,204,204}
\definecolor{orchid227119194}{RGB}{227,119,194}
\definecolor{steelblue31119180}{RGB}{31,119,180}

\begin{groupplot}[group style={group size=1 by 1}]
\nextgroupplot[
legend cell align={left},
legend style={
    fill opacity=0.8,
    draw opacity=1,
    text opacity=1,
    at={(0.98,0.40)},  
    anchor=north east,
    draw=black!20
},
tick align=outside,
tick pos=left,
x grid style={darkgray176},
xlabel={ Mean Distance (m)},
xmajorgrids,
xmin=0.05, xmax=0.25,
xtick={0.05, 0.10, 0.15, 0.20, 0.25},
xticklabels={0.05, 0.10, 0.15, 0.20, 0.25},
xtick style={color=black},
y grid style={darkgray176},
ylabel={Cumulative Probability},
ymajorgrids,
ymin=0, ymax=1.00,
ytick style={color=black},
ytick={0,0.2,0.4,0.6,0.8,1},
yticklabels={0,0.2,0.4,0.6,0.8,1}
]
\addplot [thick, crimson2143940, mark=*, mark size=1.5, mark options={solid}]
table{%
0.096689925845588 0.0256410256410256
0.107752511856508 0.0512820512820513
0.108209962653182 0.0769230769230769
0.109737713574362 0.102564102564103
0.111077818960259 0.128205128205128
0.118539665586453 0.153846153846154
0.12055304552537 0.179487179487179
0.120671590844515 0.205128205128205
0.121399445110568 0.230769230769231
0.121498096161211 0.256410256410256
0.121769404776999 0.282051282051282
0.125481442756499 0.307692307692308
0.127437045632229 0.333333333333333
0.130164201911814 0.358974358974359
0.130195668349832 0.384615384615385
0.130865790985429 0.41025641025641
0.132494762042355 0.435897435897436
0.132711138725779 0.461538461538462
0.134418012804091 0.487179487179487
0.135968167013719 0.512820512820513
0.136002114947919 0.538461538461538
0.138176127938616 0.564102564102564
0.138393659522754 0.58974358974359
0.14063177493837 0.615384615384615
0.143720063027111 0.641025641025641
0.144261297558906 0.666666666666667
0.146881041912505 0.692307692307692
0.147801089959449 0.717948717948718
0.147829601356104 0.743589743589744
0.152119274817677 0.769230769230769
0.155054918831449 0.794871794871795
0.156597035986743 0.82051282051282
0.167348409845994 0.846153846153846
0.172020828263851 0.871794871794872
0.17375618895302 0.897435897435897
0.175101329421144 0.923076923076923
0.195319457562957 0.948717948717949
0.25 1
};
\addlegendentry{MMWi-Fi}
\addplot [thick, steelblue31119180, mark=x, mark size=1.5, mark options={solid}]
table {%
0.0950948521898497 0.0263157894736842
0.095278535771836 0.0526315789473684
0.0984026690258367 0.0789473684210526
0.0987720551057729 0.105263157894737
0.0996838703572108 0.131578947368421
0.100307719049821 0.157894736842105
0.100966570814266 0.184210526315789
0.101357045271313 0.210526315789474
0.101400164556076 0.236842105263158
0.101506248958152 0.263157894736842
0.102595411249349 0.289473684210526
0.10267593315864 0.315789473684211
0.103590347220302 0.342105263157895
0.10360875589411 0.368421052631579
0.103719059329685 0.394736842105263
0.104214305608398 0.421052631578947
0.104524104982474 0.447368421052632
0.104573193076976 0.473684210526316
0.104775893432456 0.5
0.105790722703008 0.526315789473684
0.106522109031854 0.552631578947368
0.106540720451107 0.578947368421053
0.106727246868748 0.605263157894737
0.10683128627 0.631578947368421
0.106988605989796 0.657894736842105
0.107421668971007 0.684210526315789
0.107510827250928 0.710526315789474
0.10902591507451 0.736842105263158
0.109082896382614 0.763157894736842
0.109202760863235 0.789473684210526
0.10943256786822 0.815789473684211
0.110974154768469 0.842105263157895
0.111559262828951 0.868421052631579
0.115498624085327 0.894736842105263
0.117704566023303 0.921052631578947
0.120483588529677 0.947368421052632
0.121318130861377 0.973684210526316
0.134911253475156 1
};
\addlegendentry{Radar}
\addplot [thick, orchid227119194, mark=diamond, mark size=1.5, mark options={solid}]
table {%
0.1667380376872 0.0256410256410256
0.171163985887985 0.0512820512820513
0.173482085424962 0.0769230769230769
0.173855889198751 0.102564102564103
0.174435052628133 0.128205128205128
0.176305254860419 0.153846153846154
0.176852852727819 0.179487179487179
0.178579730508013 0.205128205128205
0.17919117010771 0.230769230769231
0.179861643232162 0.256410256410256
0.179928097026581 0.282051282051282
0.179984380044811 0.307692307692308
0.180705419152684 0.333333333333333
0.180796859885613 0.358974358974359
0.183246809740284 0.384615384615385
0.18380774996521 0.41025641025641
0.183995715040302 0.435897435897436
0.184164385113122 0.461538461538462
0.184695279941457 0.487179487179487
0.184728537388513 0.512820512820513
0.185740229645601 0.538461538461538
0.185803806747214 0.564102564102564
0.185956390554158 0.58974358974359
0.186086100876851 0.615384615384615
0.186119533119535 0.641025641025641
0.186417950282733 0.666666666666667
0.187874704138516 0.692307692307692
0.187878147142706 0.717948717948718
0.188259751695376 0.743589743589744
0.190060634058652 0.769230769230769
0.190245278231169 0.794871794871795
0.190432995448207 0.82051282051282
0.190501504179626 0.846153846153846
0.190590290976757 0.871794871794872
0.192021394630849 0.897435897435897
0.193049396470694 0.923076923076923
0.194282321088616 0.948717948717949
0.197713248886804 0.974358974358974
0.198115500209004 1
};
\addlegendentry{Wi-Fi}
\end{groupplot}
\end{tikzpicture}}
	}
	\hfill
	\subfloat[\label{fig:square_error}]{
		\resizebox{0.21\textwidth}{!}{
\begin{tikzpicture}

\definecolor{crimson2143940}{RGB}{214,39,40}
\definecolor{darkgray176}{RGB}{176,176,176}
\definecolor{darkturquoise23190207}{RGB}{23,190,207}
\definecolor{lightgray204}{RGB}{204,204,204}
\definecolor{orchid227119194}{RGB}{227,119,194}
\definecolor{steelblue31119180}{RGB}{31,119,180}

\begin{groupplot}[group style={group size=1 by 1}]
\nextgroupplot[
legend cell align={left},
legend style={
    fill opacity=0.8,
    draw opacity=1,
    text opacity=1,
    at={(0.98,0.40)},  
    anchor=north east,
    draw=black!20
},
tick align=outside,
tick pos=left,
x grid style={darkgray176},
xlabel={ Mean Distance (m)},
xmajorgrids,
xmin=0.05, xmax=0.25,
xtick={0.05, 0.10, 0.15, 0.20, 0.25},
xticklabels={0.05, 0.10, 0.15, 0.20, 0.25},
xtick style={color=black},
y grid style={darkgray176},
ylabel={Cumulative Probability},
ymajorgrids,
ymin=0, ymax=1.00,
ytick style={color=black},
ytick={0,0.2,0.4,0.6,0.8,1},
yticklabels={0,0.2,0.4,0.6,0.8,1}
]
\addplot [thick, crimson2143940, mark=*, mark size=1.5, mark options={solid}]
table{%
0.151284806250176 0.024390243902439
0.153373471609308 0.0487804878048781
0.156021960128572 0.0731707317073171
0.156878319958374 0.0975609756097561
0.157476426617887 0.121951219512195
0.162694903125528 0.146341463414634
0.163791819216406 0.170731707317073
0.165026490960384 0.195121951219512
0.166208215058384 0.219512195121951
0.166613383202121 0.24390243902439
0.167345843074355 0.268292682926829
0.16793515973107 0.292682926829268
0.169486844894411 0.317073170731707
0.170471888125209 0.341463414634146
0.170876894422542 0.365853658536585
0.174056878388759 0.390243902439024
0.174905364777881 0.414634146341463
0.175047264661623 0.439024390243902
0.175067912041105 0.463414634146341
0.176198684959654 0.48780487804878
0.176774195845159 0.51219512195122
0.17699151398308 0.536585365853659
0.17721297409587 0.560975609756098
0.178943790147551 0.585365853658537
0.179148256354175 0.609756097560976
0.180305022696342 0.634146341463415
0.181182770405085 0.658536585365854
0.181737093869856 0.682926829268293
0.181759367973601 0.707317073170732
0.182820860081098 0.731707317073171
0.182997289367273 0.75609756097561
0.188652625607008 0.780487804878049
0.189978740979833 0.804878048780488
0.190161917142355 0.829268292682927
0.193370954929638 0.853658536585366
0.193497114768742 0.878048780487805
0.193761065663365 0.902439024390244
0.195183632483473 0.926829268292683
0.200237142661016 0.951219512195122
0.227290565698067 0.975609756097561
0.246981171269705 1
};
\addlegendentry{MMWi-Fi}
\addplot [thick, steelblue31119180, mark=x, mark size=1.5, mark options={solid}]
table {%
0.119138227140649 0.024390243902439
0.119630342411061 0.0487804878048781
0.120886452260992 0.0731707317073171
0.121275177944966 0.0975609756097561
0.122336049803758 0.121951219512195
0.123035477386587 0.146341463414634
0.124868181680777 0.170731707317073
0.126371927706136 0.195121951219512
0.126465731027678 0.219512195121951
0.12702942018666 0.24390243902439
0.128616646588004 0.268292682926829
0.128689177441503 0.292682926829268
0.128953936609719 0.317073170731707
0.129334526809161 0.341463414634146
0.129828192010794 0.365853658536585
0.130695640310096 0.390243902439024
0.132384956163568 0.414634146341463
0.13295121918464 0.439024390243902
0.133016338502469 0.463414634146341
0.133186731592294 0.48780487804878
0.133601163190847 0.51219512195122
0.133756771311605 0.536585365853659
0.134462672221351 0.560975609756098
0.135947384567611 0.585365853658537
0.136493825663708 0.609756097560976
0.138364537346072 0.634146341463415
0.138714221120992 0.658536585365854
0.139140471232021 0.682926829268293
0.139746500204464 0.707317073170732
0.140040281074084 0.731707317073171
0.140276472984429 0.75609756097561
0.140848040152131 0.780487804878049
0.142480283130051 0.804878048780488
0.143966378341272 0.829268292682927
0.145580372992281 0.853658536585366
0.147661307988848 0.878048780487805
0.14798783352199 0.902439024390244
0.148160661847805 0.926829268292683
0.148701585645036 0.951219512195122
0.148801280499619 0.975609756097561
0.151008154693202 1
};
\addlegendentry{Radar}
\addplot [thick, orchid227119194, mark=diamond, mark size=1.5, mark options={solid}]
table {%
0.177510677707257 0.024390243902439
0.179388760861604 0.0487804878048781
0.184596447578573 0.0731707317073171
0.185121588953319 0.0975609756097561
0.188881936753309 0.121951219512195
0.189296945462925 0.146341463414634
0.189753126539669 0.170731707317073
0.190737461810013 0.195121951219512
0.190795036509304 0.219512195121951
0.192142259475388 0.24390243902439
0.192579599490044 0.268292682926829
0.192719946179266 0.292682926829268
0.193123739434428 0.317073170731707
0.19358026009212 0.341463414634146
0.193677259970493 0.365853658536585
0.194281573151784 0.390243902439024
0.194925402168853 0.414634146341463
0.195175061689038 0.439024390243902
0.195559134664652 0.463414634146341
0.195728095943846 0.48780487804878
0.195923652609814 0.51219512195122
0.197291601770641 0.536585365853659
0.197437714693878 0.560975609756098
0.197511246494567 0.585365853658537
0.197743219274523 0.609756097560976
0.198084758908326 0.634146341463415
0.199882217768385 0.658536585365854
0.200400111328957 0.682926829268293
0.200694151640717 0.707317073170732
0.201105741370882 0.731707317073171
0.201858278852905 0.75609756097561
0.203382095885263 0.780487804878049
0.204220107192983 0.804878048780488
0.205650404469255 0.829268292682927
0.207022750984522 0.853658536585366
0.209373625362996 0.878048780487805
0.210893726270434 0.902439024390244
0.211635224909913 0.926829268292683
0.214846203492477 0.951219512195122
0.215907290771855 0.975609756097561
0.240462100313351 1
};
\addlegendentry{Wi-Fi}
\end{groupplot}
\end{tikzpicture}}
	}
	\subfloat[\label{fig:zshape_error}]{
		\resizebox{0.21\textwidth}{!}{
\begin{tikzpicture}

\definecolor{crimson2143940}{RGB}{214,39,40}
\definecolor{darkgray176}{RGB}{176,176,176}
\definecolor{darkturquoise23190207}{RGB}{23,190,207}
\definecolor{lightgray204}{RGB}{204,204,204}
\definecolor{orchid227119194}{RGB}{227,119,194}
\definecolor{steelblue31119180}{RGB}{31,119,180}

\begin{groupplot}[group style={group size=1 by 1}]
\nextgroupplot[
legend cell align={left},
legend style={
    fill opacity=0.8,
    draw opacity=1,
    text opacity=1,
    at={(0.98,0.40)},  
    anchor=north east,
    draw=black!20
},
tick align=outside,
tick pos=left,
x grid style={darkgray176},
xlabel={ Mean Distance (m)},
xmajorgrids,
xmin=0.05, xmax=0.25,
xtick={0.05, 0.10, 0.15, 0.20, 0.25},
xticklabels={0.05, 0.10, 0.15, 0.20, 0.25},
xtick style={color=black},
y grid style={darkgray176},
ylabel={Cumulative Probability},
ymajorgrids,
ymin=0, ymax=1.00,
ytick style={color=black},
ytick={0,0.2,0.4,0.6,0.8,1},
yticklabels={0,0.2,0.4,0.6,0.8,1}
]
\addplot [thick, crimson2143940, mark=*, mark size=1.5, mark options={solid}]
table{%
0.123680715915541 0.0263157894736842
0.126084957270268 0.0526315789473684
0.129405673399702 0.0789473684210526
0.136313987470527 0.105263157894737
0.142838261419582 0.131578947368421
0.143538474393597 0.157894736842105
0.143678924759477 0.184210526315789
0.144140086573216 0.210526315789474
0.146189808712797 0.236842105263158
0.146542948114921 0.263157894736842
0.147870280323443 0.289473684210526
0.148210027337976 0.315789473684211
0.14855207034351 0.342105263157895
0.148771131803659 0.368421052631579
0.148853001595969 0.394736842105263
0.151082201169372 0.421052631578947
0.151813107019654 0.447368421052632
0.154744657832246 0.473684210526316
0.157296058881104 0.5
0.157830381299983 0.526315789473684
0.15822974017896 0.552631578947368
0.158739666188963 0.578947368421053
0.164844499642895 0.605263157894737
0.165297156824636 0.631578947368421
0.166323152643888 0.657894736842105
0.167387064159207 0.684210526315789
0.167892842113222 0.710526315789474
0.168159277886601 0.736842105263158
0.169754035129484 0.763157894736842
0.171610229029177 0.789473684210526
0.1759093328395 0.815789473684211
0.17644565854712 0.842105263157895
0.176592819822418 0.868421052631579
0.180258380209265 0.894736842105263
0.18043070077182 0.921052631578947
0.182488850382267 0.947368421052632
0.185312319916279 0.973684210526316
0.203748674807778 1
};
\addlegendentry{MMWi-Fi}
\addplot [thick, steelblue31119180, mark=x, mark size=1.5, mark options={solid}]
table {%
0.114649120253334 0.0263157894736842
0.114715850984134 0.0526315789473684
0.118439688633105 0.0789473684210526
0.11872954376711 0.105263157894737
0.119591584590106 0.131578947368421
0.125525095972269 0.157894736842105
0.125960523999245 0.184210526315789
0.125962010257182 0.210526315789474
0.127069805453912 0.236842105263158
0.128394317192228 0.263157894736842
0.130381071940991 0.289473684210526
0.133903159013766 0.315789473684211
0.135184766570532 0.342105263157895
0.135866577365621 0.368421052631579
0.135908650047824 0.394736842105263
0.137411933423207 0.421052631578947
0.137517961723304 0.447368421052632
0.138060591594227 0.473684210526316
0.141612583272761 0.5
0.141841605407533 0.526315789473684
0.142934427788308 0.552631578947368
0.143224783181384 0.578947368421053
0.143810330398107 0.605263157894737
0.143998121297453 0.631578947368421
0.14424495586789 0.657894736842105
0.148157254425459 0.684210526315789
0.151574420763013 0.710526315789474
0.151930014802604 0.736842105263158
0.156329599958005 0.763157894736842
0.161034036152776 0.789473684210526
0.162959213622843 0.815789473684211
0.169849610027841 0.842105263157895
0.173372920734437 0.868421052631579
0.176241348508569 0.894736842105263
0.181052847917692 0.921052631578947
0.184561567197525 0.947368421052632
0.188736041360021 0.973684210526316
0.191487655234039 1
};
\addlegendentry{Radar}
\addplot [thick, orchid227119194, mark=diamond, mark size=1.5, mark options={solid}]
table {%
0.140206939903114 0.0263157894736842
0.150415765034726 0.0526315789473684
0.156156936463862 0.0789473684210526
0.159092926667896 0.105263157894737
0.160520817965082 0.131578947368421
0.160524998088163 0.157894736842105
0.161260709261468 0.184210526315789
0.163705399611951 0.210526315789474
0.16415857915497 0.236842105263158
0.164343567810913 0.263157894736842
0.166608659850356 0.289473684210526
0.167078422027868 0.315789473684211
0.167119567337854 0.342105263157895
0.168895935679567 0.368421052631579
0.169607122243712 0.394736842105263
0.170268717596166 0.421052631578947
0.171615812093156 0.447368421052632
0.17214563978498 0.473684210526316
0.173351799069294 0.5
0.173602570449048 0.526315789473684
0.175521739908643 0.552631578947368
0.177056252121737 0.578947368421053
0.178513824808139 0.605263157894737
0.17864600670795 0.631578947368421
0.179361820075177 0.657894736842105
0.180191697048047 0.684210526315789
0.180351460359553 0.710526315789474
0.181144702951097 0.736842105263158
0.18398819359773 0.763157894736842
0.185603821747455 0.789473684210526
0.190209096407697 0.815789473684211
0.196362975554313 0.842105263157895
0.197752305070331 0.868421052631579
0.197884034461195 0.894736842105263
0.198019183384278 0.921052631578947
0.236968945294118 0.947368421052632
0.24122571 0.973684210526316
0.25 1
};
\addlegendentry{Wi-Fi}
\end{groupplot}
\end{tikzpicture}}
	}
	\hfill
	\subfloat[\label{fig:rectangle_error}]{
		\resizebox{0.21\textwidth}{!}{
\begin{tikzpicture}

\definecolor{crimson2143940}{RGB}{214,39,40}
\definecolor{darkgray176}{RGB}{176,176,176}
\definecolor{darkturquoise23190207}{RGB}{23,190,207}
\definecolor{lightgray204}{RGB}{204,204,204}
\definecolor{orchid227119194}{RGB}{227,119,194}
\definecolor{steelblue31119180}{RGB}{31,119,180}

\begin{groupplot}[group style={group size=1 by 1}]
\nextgroupplot[
legend cell align={left},
legend style={
    fill opacity=0.8,
    draw opacity=1,
    text opacity=1,
    at={(0.98,0.40)},  
    anchor=north east,
    draw=black!20
},
tick align=outside,
tick pos=left,
x grid style={darkgray176},
xlabel={ Mean Distance (m)},
xmajorgrids,
xmin=0.05, xmax=0.25,
xtick={0.05, 0.10, 0.15, 0.20, 0.25},
xticklabels={0.05, 0.10, 0.15, 0.20, 0.25},
xtick style={color=black},
y grid style={darkgray176},
ylabel={Cumulative Probability},
ymajorgrids,
ymin=0, ymax=1.00,
ytick style={color=black},
ytick={0,0.2,0.4,0.6,0.8,1},
yticklabels={0,0.2,0.4,0.6,0.8,1}
]
\addplot [thick, crimson2143940, mark=*, mark size=1.5, mark options={solid}]
table{%
0.0831264464593243 0.025
0.0964785135747264 0.05
0.100772438008336 0.075
0.10936864464033 0.1
0.111532090008356 0.125
0.119782800651376 0.15
0.125983987598261 0.175
0.127232818635411 0.2
0.135387355679051 0.225
0.135635817039042 0.25
0.147517045102548 0.275
0.147829554513426 0.3
0.149445096149822 0.325
0.14982939944356 0.35
0.150737450878495 0.375
0.153223239935073 0.4
0.153373516668659 0.425
0.157588011222374 0.45
0.158562169112874 0.475
0.159065613334034 0.5
0.163524130328881 0.525
0.164709011205719 0.55
0.165171239146263 0.575
0.170829261329089 0.6
0.177119897220813 0.625
0.177414165372143 0.65
0.183293005029974 0.675
0.183301740757679 0.7
0.188011492687596 0.725
0.188431069657103 0.75
0.193753322983979 0.775
0.205999347411003 0.8
0.210988009404494 0.825
0.213026430974484 0.85
0.21477643426688 0.875
0.217061941200437 0.9
0.218382661502661 0.925
0.226030965420069 0.95
0.237609353498981 0.975
0.245319700555532 1
};
\addlegendentry{MMWi-Fi}
\addplot [thick, steelblue31119180, mark=x, mark size=1.5, mark options={solid}]
table {%
0.115071170812695 0.025
0.122380434291409 0.05
0.124918494357544 0.075
0.13034170376253 0.1
0.131534677917007 0.125
0.133447279873244 0.15
0.135520416477389 0.175
0.136103585330792 0.2
0.136273318776152 0.225
0.13756259086898 0.25
0.137668354559538 0.275
0.138839360629929 0.3
0.139305043041823 0.325
0.140659792563594 0.35
0.141228757368528 0.375
0.1415427225423 0.4
0.143474803050847 0.425
0.143689717069686 0.45
0.144071199487935 0.475
0.144113933435102 0.5
0.144518524739543 0.525
0.145196080822565 0.55
0.14543552501409 0.575
0.148260049867712 0.6
0.148576281055975 0.625
0.149635930463523 0.65
0.150359099713307 0.675
0.15142917770233 0.7
0.153900022222869 0.725
0.158227813299386 0.75
0.16044383875346 0.775
0.164920272393045 0.8
0.166852403833397 0.825
0.167761367340401 0.85
0.180738862415546 0.875
0.183099814467169 0.9
0.18376963782426 0.925
0.190644746726138 0.95
0.201750577270302 0.975
0.222930349707602 1
};
\addlegendentry{Radar}
\addplot [thick, orchid227119194, mark=diamond, mark size=1.5, mark options={solid}]
table {%
0.125216351428571 0.025
0.130190996980275 0.05
0.130676235226725 0.075
0.132488210612245 0.1
0.137060612996255 0.125
0.138363805859448 0.15
0.141632886358188 0.175
0.143356193829787 0.2
0.153051246071429 0.225
0.159051377732824 0.25
0.162250219433327 0.275
0.16459616474005 0.3
0.16534302875 0.325
0.171104432979616 0.35
0.174312597124608 0.375
0.174344251818182 0.4
0.177447092466425 0.425
0.178902528235294 0.45
0.18168451 0.475
0.182110965384615 0.5
0.185214288732211 0.525
0.186342441842105 0.55
0.189533273333333 0.575
0.190952210035643 0.6
0.191007534 0.625
0.191009414444444 0.65
0.191344087362949 0.675
0.192105734369513 0.7
0.192785754850476 0.725
0.193024331388554 0.75
0.19443350047619 0.775
0.1949316875 0.8
0.195548563533108 0.825
0.198800099736842 0.85
0.199522673055556 0.875
0.199871961411209 0.9
0.200422381379815 0.925
0.205538373863636 0.95
0.205730854285714 0.975
0.207938798891357 1
};
\addlegendentry{Wi-Fi}
\end{groupplot}
\end{tikzpicture}}
	}
	\subfloat[\label{fig:lhour_error}]{
		\resizebox{0.21\textwidth}{!}{
\begin{tikzpicture}

\definecolor{crimson2143940}{RGB}{214,39,40}
\definecolor{darkgray176}{RGB}{176,176,176}
\definecolor{darkturquoise23190207}{RGB}{23,190,207}
\definecolor{lightgray204}{RGB}{204,204,204}
\definecolor{orchid227119194}{RGB}{227,119,194}
\definecolor{steelblue31119180}{RGB}{31,119,180}

\begin{groupplot}[group style={group size=1 by 1}]
\nextgroupplot[
legend cell align={left},
legend style={
    fill opacity=0.8,
    draw opacity=1,
    text opacity=1,
    at={(0.98,0.40)},  
    anchor=north east,
    draw=black!20
},
tick align=outside,
tick pos=left,
x grid style={darkgray176},
xlabel={ Mean Distance (m)},
xmajorgrids,
xmin=0.05, xmax=0.25,
xtick={0.05, 0.10, 0.15, 0.20, 0.25},
xticklabels={0.05, 0.10, 0.15, 0.20, 0.25},
xtick style={color=black},
y grid style={darkgray176},
ylabel={Cumulative Probability},
ymajorgrids,
ymin=0, ymax=1.00,
ytick style={color=black},
ytick={0,0.2,0.4,0.6,0.8,1},
yticklabels={0,0.2,0.4,0.6,0.8,1}
]
\addplot [thick, crimson2143940, mark=*, mark size=1.5, mark options={solid}]
table{%
0.142068852347947 0.0256410256410256
0.14484112684299 0.0512820512820513
0.145667293343541 0.0769230769230769
0.146641974620205 0.102564102564103
0.148252265351324 0.128205128205128
0.148620983785721 0.153846153846154
0.150298615004925 0.179487179487179
0.157309429074814 0.205128205128205
0.159327872171462 0.230769230769231
0.159593368641235 0.256410256410256
0.162278398158244 0.282051282051282
0.165914648083868 0.307692307692308
0.16849682387456 0.333333333333333
0.169120192333216 0.358974358974359
0.171189699484585 0.384615384615385
0.17540257459302 0.41025641025641
0.177193167684185 0.435897435897436
0.177352478697067 0.461538461538462
0.177993875559607 0.487179487179487
0.178384085295023 0.512820512820513
0.180601181386995 0.538461538461538
0.183044010575258 0.564102564102564
0.189690002770652 0.58974358974359
0.189855437200285 0.615384615384615
0.1906237408375 0.641025641025641
0.190631914285473 0.666666666666667
0.195133646636309 0.692307692307692
0.197147550811864 0.717948717948718
0.200011012160451 0.743589743589744
0.202203751466882 0.769230769230769
0.206409075592761 0.794871794871795
0.206498623222508 0.82051282051282
0.208507917088417 0.846153846153846
0.208815591843378 0.871794871794872
0.212690339641789 0.897435897435897
0.21649336175375 0.923076923076923
0.217480084753022 0.948717948717949
0.240299034386048 0.974358974358974
0.25 1
};
\addlegendentry{MMWi-Fi}
\addplot [thick, steelblue31119180, mark=x, mark size=1.5, mark options={solid}]
table {%
0.0995847464250234 0.0256410256410256
0.10419491490039 0.0512820512820513
0.104350943953348 0.0769230769230769
0.106869302377724 0.102564102564103
0.10690451271771 0.128205128205128
0.108208074659682 0.153846153846154
0.109842046425045 0.179487179487179
0.1107927937453 0.205128205128205
0.116625915032241 0.230769230769231
0.117545755412997 0.256410256410256
0.118769091702417 0.282051282051282
0.121086815152765 0.307692307692308
0.121276587184846 0.333333333333333
0.121494881174108 0.358974358974359
0.125274409613968 0.384615384615385
0.125735966283636 0.41025641025641
0.126803723910677 0.435897435897436
0.130836268912371 0.461538461538462
0.131362153763851 0.487179487179487
0.134519674388755 0.512820512820513
0.135441704638144 0.538461538461538
0.137692877464938 0.564102564102564
0.138430385596947 0.58974358974359
0.141743485440178 0.615384615384615
0.144879139175139 0.641025641025641
0.146323485472488 0.666666666666667
0.151218518691844 0.692307692307692
0.153258926030843 0.717948717948718
0.156389242210238 0.743589743589744
0.160225024709674 0.769230769230769
0.169715356915106 0.794871794871795
0.172368392125767 0.82051282051282
0.172778233433391 0.846153846153846
0.181103301076971 0.871794871794872
0.182853677760853 0.897435897435897
0.186298219043604 0.923076923076923
0.186469454540296 0.948717948717949
0.187400976271446 0.974358974358974
0.188695686058868 1
};
\addlegendentry{Radar}
\addplot [thick, orchid227119194, mark=diamond, mark size=1.5, mark options={solid}]
table {%
0.137431056666667 0.0256410256410256
0.166082451428571 0.0512820512820513
0.182454845874509 0.0769230769230769
0.187460694142869 0.102564102564103
0.188943547707851 0.128205128205128
0.188991339958272 0.153846153846154
0.190047643015079 0.179487179487179
0.191049993597792 0.205128205128205
0.191476816882277 0.230769230769231
0.192377754776922 0.256410256410256
0.192802849286199 0.282051282051282
0.193561192263187 0.307692307692308
0.193795412648496 0.333333333333333
0.19463308223715 0.358974358974359
0.194881279641754 0.384615384615385
0.195267665117361 0.41025641025641
0.197787848844593 0.435897435897436
0.198250052552241 0.461538461538462
0.198396142817233 0.487179487179487
0.198439417353581 0.512820512820513
0.200955500490789 0.538461538461538
0.201045631614688 0.564102564102564
0.201099562861498 0.58974358974359
0.201169602367551 0.615384615384615
0.201218382566672 0.641025641025641
0.202034671229448 0.666666666666667
0.202610737488391 0.692307692307692
0.209771333480521 0.717948717948718
0.210819933168682 0.743589743589744
0.213144806022262 0.769230769230769
0.226273975 0.794871794871795
0.236083212 0.82051282051282
0.236861766153846 0.846153846153846
0.239616006153846 0.871794871794872
0.241923207692308 0.897435897435897
0.25 0.923076923076923
0.25 0.948717948717949
0.25 0.974358974358974
0.25 1
};
\addlegendentry{Wi-Fi}
\end{groupplot}
\end{tikzpicture}}
	}
\caption{The localization error for different moving patterns with sensors. (a) Hourglass Shape (b) Diamond Shape (c) Square Shape (d) Z Shape (e) Rectangle Shape (f) Larger Hourglass Shape.}
\label{fig:figure_4}
\end{figure}
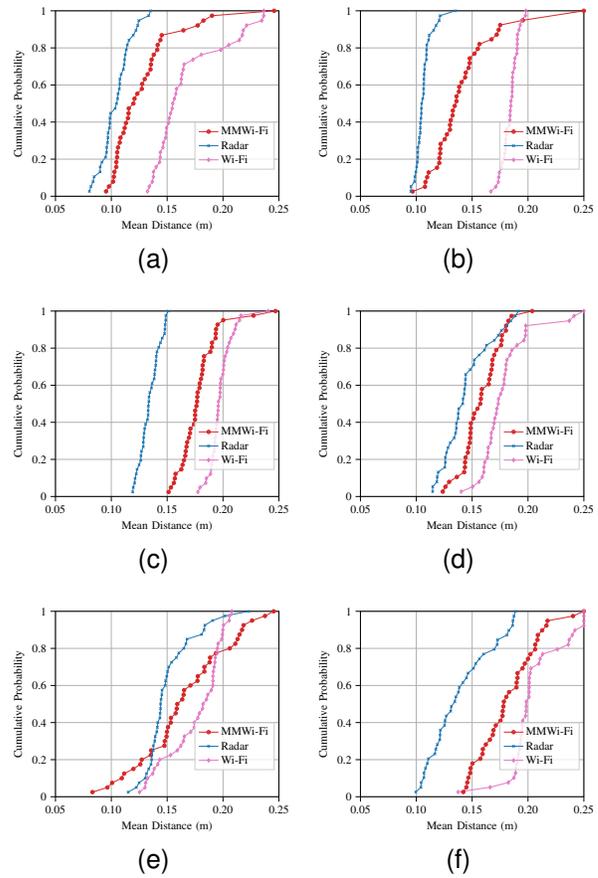


Quantitative performance analysis, shown in Figure \ref{fig:figure_4}, reveals several key insights regarding system accuracy. Localization accuracy is quantified by computing the error relative to ground-truth positions. At each timestamp, we identify the points that fall within a 0.25m radius of the ground-truth position. The localization error is then calculated as the average Euclidean distance between these points and the ground-truth. If no points are found within this threshold, a maximum error of 0.25m is assigned for that timestamp. This evaluation methodology aligns with the metrics employed in \cite{wuWiTrajRobustIndoor2021}, facilitating direct performance comparisons. The Cumulative Distribution Function (CDF) plots in Figure \ref{fig:figure_4} represent the average error computed across multiple recording sessions, with each motion pattern tested 40 times.

\begin{table}[ht!]
	\caption{Movement Pattern Performance Analysis}
	\vspace{0.5em}\centering
	\begin{tabular}{l cc cc}
		\hline
        Pattern & \multicolumn{2}{c}{Distance Error(m)$\downarrow$} & \multicolumn{2}{c}{Association Rate$\uparrow$} \\
        & Mean & Median & Mean & Median \\
		\hline
        \multicolumn{5}{l}{\textbf{MMWi-Fi(MMWiLoc)}} \\
		Hourglass & 0.125 & 0.121 & \textbf{0.77} & \textbf{0.77} \\
        Diamond & 0.138 & 0.136 & 0.74 & \textbf{0.76} \\
        Square & 0.165 & 0.159 & \textbf{0.70} & \textbf{0.70} \\
        Z-shape & 0.160 & 0.157 & \textbf{0.60} & \textbf{0.60} \\
        Rectangle & 0.166 & 0.164 & \textbf{0.52} & \textbf{0.52} \\
        Large Hourglass & 0.182 & 0.180 & \textbf{0.62} & \textbf{0.62} \\
        \hline
        \multicolumn{5}{l}{\textbf{Radar}} \\
        Hourglass & \textbf{0.115} & \textbf{0.113} & 0.74 & 0.74 \\
		Diamond & \textbf{0.112} & \textbf{0.110} & \textbf{0.75} & 0.75 \\
        Square & \textbf{0.125} & \textbf{0.123} & 0.51 & 0.51 \\
        Z-shape & \textbf{0.130} & \textbf{0.128} & 0.57 & 0.57 \\
        Rectangle & \textbf{0.147} & \textbf{0.145} & 0.40 & 0.40 \\
        Large Hourglass & \textbf{0.143} & \textbf{0.141} & 0.56 & 0.56 \\
        \hline
        \multicolumn{5}{l}{\textbf{2.4GHz Wi-Fi(Widar2)}} \\
        Hourglass & 0.177 & 0.175 & 0.45 & 0.45 \\
		Diamond & 0.183 & 0.181 & 0.48 & 0.48 \\
        Square & 0.180 & 0.178 & 0.45 & 0.45 \\
        Z-shape & 0.185 & 0.183 & 0.42 & 0.42 \\
        Rectangle & 0.182 & 0.180 & 0.38 & 0.38 \\
        Large Hourglass & 0.212 & 0.210 & 0.35 & 0.35 \\
		\hline
	\end{tabular}
	\label{table_movement}
\end{table}

The association rate is defined as the ratio of points that fall within the 0.25m threshold relative to all detected points. As summarized in Table \ref{table_movement} with best performance metrics highlighted in \textbf{bold}, MMWiLoc demonstrates strong overall performance across various movement patterns, particularly excelling in simpler trajectories. For basic patterns like hourglass and diamond shapes, MMWiLoc achieves performance closely matching dedicated radar systems while significantly outperforming 2.4GHz Wi-Fi approaches. Performance gradually declines with increasing pattern complexity and range, as evidenced by the rectangle and large hourglass patterns, though MMWiLoc consistently maintains superior association rates compared to 2.4GHz Wi-Fi. The radar system shows the most consistent distance accuracy across all patterns but exhibits notable association rate drops in complex scenarios. These results demonstrate that while all systems experience performance degradation with increased pattern complexity and range, MMWiLoc maintains competitive performance against specialized radar systems and consistently outperforms traditional Wi-Fi approaches.

\subsection{Algorithm Analysis}
\label{ablation}

\begin{figure}[htp]
	\centering
	\subfloat[\label{fig:secter_pattern_0}]{
		\resizebox{0.149\textwidth}{!}{\begin{tikzpicture}

\pgfplotstableread[col sep=comma]{
22.047779083251953,17.595255037502586
22.152507781982422,18.160644778124656
22.178956985473633,18.239833632378957
23.215679168701172,18.10733293985708
24.072484970092773,18.0232904957717
24.543468475341797,18.31238560242469
24.775253295898438,18.623532687954036
25.171308517456055,18.561735914148315
25.908735275268555,18.776006195925618
25.784832000732422,18.935279095492252
26.367712020874023,19.22425096171758
26.249197006225586,18.97897830916921
26.85995101928711,19.372413379119834
27.006420135498047,19.15329514655333
27.250225067138672,19.797999771016684
27.227046966552734,19.190722193346918
27.287506103515625,19.492448864609194
27.308448791503906,19.8055608352923
27.741060256958008,19.74842074337336
28.010318756103516,20.78977607556686
28.634788513183594,21.39189122841685
28.801483154296875,23.348756179991398
29.41189193725586,24.07605730913776
30.12466812133789,25.592701429614834
30.367969512939453,26.927667734820382
30.656993865966797,28.271260572977077
30.38671112060547,29.054212834797042
31.470335006713867,29.952998624071515
31.750581741333008,30.190959557020317
32.06251907348633,30.034713934158304
31.82044219970703,30.834055419704214
31.815719604492188,31.515768895016226
32.358062744140625,31.93119787558991
32.43145751953125,33.01739882930726
32.995399475097656,33.46169559223321
32.95810317993164,33.31449829862112
33.058937072753906,33.16011382444068
32.83601379394531,33.05698313924537
33.06276321411133,32.21091025102275
33.06398391723633,31.70549089772877
33.29032516479492,30.584437774944174
33.15053939819336,30.755242564923115
33.38393783569336,30.78274570555855
33.03336715698242,30.861181143182485
32.94729995727539,31.298037750122877
32.69707107543945,31.75337344888248
32.45424270629883,32.49753196120754
32.62394714355469,33.082705711704705
32.173789978027344,33.09194122043362
32.6378173828125,33.36691735116915
32.202762603759766,33.58053965565607
32.53429412841797,32.54314899701503
32.196144104003906,32.66862290428755
32.32830047607422,31.48622032138812
32.87752914428711,31.93891292459579
33.13140869140625,31.654882650343662
33.76031494140625,31.94704726218344
33.5159797668457,32.427547215558135
33.890865325927734,32.58573414659623
33.777645111083984,33.312926279850466
34.45978546142578,33.618334849710685
34.80129623413086,33.822807364844444
35.293861389160156,33.88713249558677
35.1525764465332,34.150680527018125
35.19440460205078,34.56689675039571
34.92522430419922,34.61929857441454
34.938385009765625,34.73140787341525
34.75429916381836,34.768883153041614
34.768035888671875,34.14776826361084
34.100189208984375,34.31952187085221
34.379451751708984,33.83837022369872
34.08082962036133,33.46177585551779
34.1842155456543,33.000052171751804
33.48090744018555,32.814143333668014
33.206058502197266,32.796987660243666
32.96336364746094,32.469684878157224
32.51654815673828,32.66541851138032
32.49996566772461,32.24039413472889
32.444576263427734,31.978519662981103
32.435420989990234,31.915838778746533
32.30519485473633,32.15443951127523
32.107147216796875,31.845448269638815
32.38253402709961,31.787410238735273
32.43736267089844,31.661402772849705
32.5117301940918,31.767798244582824
32.68006896972656,31.89063570866065
33.0159912109375,32.23284907056163
33.42100524902344,32.68878068892572
33.24262237548828,32.22488435525561
33.175437927246094,32.492050512932565
32.940147399902344,32.40061833095631
32.766292572021484,32.468168947798716
32.54807662963867,32.299302409342154
32.508609771728516,32.420652268403025
32.56842803955078,32.385160490637254
32.44892883300781,32.3554037283282
32.397560119628906,32.50027480793021
32.33241653442383,31.761242485229243
32.49858474731445,31.365872879198502
32.31314468383789,30.498254477868805
32.25140380859375,30.423023699715955
31.985916137695312,30.151020903459834
32.04288864135742,30.261093203068782
32.38764572143555,30.458601195965564
32.759891510009766,30.43281960147263
32.79773712158203,30.679033024149405
32.16904830932617,30.31088846828048
32.01960754394531,30.36684161284866
32.11172103881836,30.201151097074824
32.097103118896484,29.70444979622243
31.896373748779297,29.121175617040578
31.542320251464844,28.433842553432896
31.389354705810547,27.618289659561505
30.967247009277344,26.730324750708093
30.718387603759766,26.21289394105154
30.73907470703125,26.25347846371018
30.671485900878906,26.67700217325926
30.634395599365234,27.547460934659295
30.51946449279785,28.161097752505377
30.676998138427734,28.928431737742088
31.156124114990234,28.45530350851525
31.426877975463867,27.97310928729081
31.149829864501953,26.927222523534333
30.95660400390625,26.017569114142947
30.670345306396484,24.377478575167316
31.038740158081055,23.239081221606778
31.074462890625,22.207690998653305
31.12995147705078,21.875585216069794
30.53707504272461,21.74878491727204
30.65999984741211,21.80388801587513
30.579004287719727,21.746327726954796
30.980484008789062,21.696337850089265
30.75640106201172,21.797154575155524
30.830177307128906,21.839976566680775
30.513620376586914,21.959148867529777
30.552169799804688,22.394405423729257
30.658004760742188,22.21724259572003
30.678085327148438,22.74224745417562
30.492935180664062,22.772249142197577
30.44074249267578,22.937919659267948
30.951011657714844,22.908710597271348
30.9754581451416,22.829921581977327
30.914531707763672,23.29206392268542
30.5953426361084,23.139453309288008
30.67926025390625,23.12456006102792
30.92879867553711,23.32753318616076
31.091259002685547,23.08376324207181
31.007381439208984,23.07547598890046
30.834369659423828,22.683501774212655
30.864286422729492,23.184053750397485
30.840423583984375,22.746112008690872
30.689781188964844,22.863971623582753
30.611265182495117,22.46270395039391
30.624055862426758,22.04622841118976
30.22064208984375,22.44917311173009
29.59246826171875,22.723570779274205
28.742145538330078,22.356513466076244
29.119462966918945,22.6542632262439
28.659608840942383,22.30505639777489
28.430274963378906,22.157218537869557
27.651042938232422,21.94207619167892
27.36762809753418,21.266055705879484
26.47503662109375,21.11450270914727
24.947494506835938,20.568918384133248
23.849170684814453,20.10937998802822
23.119144439697266,19.909576538959517
22.544628143310547,19.7842084496622
21.696250915527344,19.385041386164183
20.92166519165039,19.043773067707164
20.429697036743164,18.680959229738498
19.95161247253418,18.48419812217379
20.142227172851562,18.30637592437055
20.35434341430664,19.068170181468076
20.961973190307617,19.352169581866367
21.598934173583984,20.04758276103309
22.606307983398438,20.681893904143035
23.80276107788086,21.459686761280903
24.905838012695312,21.453450346649298
25.72366714477539,21.764967533774943
26.41348648071289,22.01411319813037
27.22406578063965,22.156683620790258
28.10698699951172,22.842214871524295
28.760791778564453,23.470792528529685
29.314998626708984,25.00582138679423
29.6732120513916,26.76087403896965
30.419048309326172,28.08197743442974
30.963285446166992,29.206715972955305
31.62775421142578,29.619749737945398
31.963491439819336,30.613481731819782
32.187644958496094,31.03152854007771
32.512046813964844,31.37428393614192
32.90359878540039,31.349572533832472
33.61854934692383,31.47275339496288
33.342018127441406,31.707825703522108
33.382911682128906,31.431669071292667
33.21448516845703,32.562460994133914
33.95210266113281,32.594208970383285
34.52915954589844,33.22303986386194
34.53433609008789,33.75790874165608
34.50431442260742,33.69035513912938
34.359439849853516,34.469755922896766
34.8068962097168,34.95996571108334
35.31736373901367,34.71945445403575
35.497657775878906,35.09689137800399
35.3930549621582,34.639448462271744
35.435340881347656,34.51328369479097
35.39849853515625,33.840327675309084
35.6290283203125,33.3887516505077
35.48282241821289,32.920641131073566
35.692466735839844,32.69154920519937
35.48405075073242,32.3950553155626
35.53217315673828,32.519413066224004
35.45620346069336,32.9979336168451
35.44514846801758,32.971070475049046
35.6062126159668,33.15810026502366
35.76640701293945,33.69841524182697
35.7652702331543,33.45369200567593
35.73735046386719,33.66812263541536
35.5788459777832,33.60071144956587
35.631690979003906,33.340937374276486
35.352325439453125,33.360152584247075
35.351585388183594,33.44825452891611
35.717254638671875,33.78832516458937
35.89787673950195,34.48581380547739
35.51944351196289,35.063851591611964
35.268653869628906,35.723057898157165
34.84796142578125,35.47602428537853
35.1628532409668,34.74882553460172
34.71097183227539,34.43365515514184
35.17047119140625,34.153813787305126
34.872554779052734,33.750279244387535
35.38045120239258,33.59063817227233
34.85674285888672,32.92481395865059
34.61498260498047,33.3063959058137
33.69785690307617,33.00732609989028
33.572628021240234,33.42576936474173
33.57661056518555,34.08296886737213
33.60420227050781,33.630029060930006
33.22093963623047,33.62058330372205
32.93465042114258,34.14821086672422
32.584022521972656,34.194278312384455
}\mytable;

\begin{polaraxis}[
    width=0.8\textwidth,
    height=0.8\textwidth,
    grid=major,
    xtick={-90,-60,-30,0,30,60,90},
    xticklabels={-90,-60,-30,0,30,60,90},
    xmin=-90,
    xmax=90,
    ymin=0,
    ymax=38,
    legend pos=outer north east,
    legend style={draw=black},
    major grid style={gray!30}
]

\addplot[blue, thick] table[
    x expr={\coordindex*180/241 - 90},
    y index=0,
] {\mytable};

\addplot[red, thick] table[
    x expr={\coordindex*180/241 - 90},
    y index=1,
] {\mytable};

\legend{Original,Revised}
\end{polaraxis}
\end{tikzpicture}}
	}
	\hfill
	\subfloat[\label{fig:secter_pattern_1}]{
		\resizebox{0.149\textwidth}{!}{\begin{tikzpicture}

\pgfplotstableread[col sep=comma]{
28.931190490722656,19.524524671381826
28.471542358398438,19.67891946426137
28.096084594726562,19.040474756777105
28.673625946044922,19.30024500020476
28.74311065673828,19.307036069961743
28.708892822265625,19.379806041565555
28.53182029724121,19.216034058567317
28.710336685180664,19.505597961549757
28.946025848388672,19.32560979649154
28.593753814697266,19.00987159005481
29.192474365234375,19.442132739559202
29.177814483642578,19.391829905613395
29.77829933166504,19.438328013768455
29.644147872924805,19.48176479148534
29.70977783203125,19.33705195922561
29.498390197753906,19.741940395312053
29.468271255493164,19.51573836073506
29.364391326904297,19.52694437049696
29.622989654541016,19.84097809702686
29.699363708496094,20.000674089059597
29.841716766357422,20.49863268713734
29.586833953857422,21.063569336736023
29.543624877929688,22.02409677644002
29.803678512573242,23.255242882575793
29.580665588378906,24.351252504962332
29.304065704345703,25.47760718403886
28.506179809570312,26.80079594281927
28.772808074951172,27.67117111109306
28.868995666503906,28.03407434240746
28.696578979492188,28.42256867866558
28.12151336669922,28.93679084374023
27.572404861450195,28.93754952887647
27.797517776489258,28.985242739890186
27.55093002319336,29.428395076370105
27.829940795898438,29.00742239456673
27.56291389465332,28.94706928754164
27.414878845214844,29.575866234600806
27.112022399902344,29.99734534299052
27.25909423828125,30.098841306299594
27.383222579956055,30.156201932229997
27.733959197998047,30.260053732035324
27.661006927490234,30.01289363827265
27.960338592529297,29.63989193965851
27.689979553222656,29.26616498513581
27.790334701538086,29.27541681413411
27.68743133544922,29.91204891199416
27.425495147705078,29.747339238198258
27.604673385620117,29.98310667553884
27.207534790039062,30.191292952832654
27.524307250976562,30.697921421904525
26.94995880126953,29.889733819187533
27.026134490966797,29.9408498056509
26.269685745239258,29.582623568645378
25.81751251220703,28.940988202146094
25.34870147705078,28.85488275531134
24.898855209350586,28.468317141394092
24.660934448242188,27.876542034187537
23.923095703125,27.715945774812095
23.474105834960938,27.16129844334474
22.146509170532227,26.623573123374896
21.309490203857422,26.087511705360356
19.919788360595703,25.947851997166453
19.364362716674805,25.419351642370273
18.4360294342041,24.99822865885249
18.125967025756836,25.16575402105498
17.77242088317871,25.335029414588828
18.350353240966797,24.773393964696826
19.237525939941406,24.525353497401778
20.515546798706055,24.043086228995882
20.946136474609375,23.914997408875752
21.986827850341797,23.27523796235452
22.00483512878418,22.985356286280304
22.72747039794922,23.311593337557134
22.777584075927734,23.882593311933718
23.38364028930664,23.92817601884341
23.833953857421875,25.17964433746021
23.825916290283203,25.750087219901353
24.395355224609375,26.442321915980664
24.657468795776367,27.310436505345237
25.21875,27.723017247702817
25.71924591064453,28.268845468628193
26.04985809326172,29.08620234765026
26.789838790893555,29.493758155993376
27.192378997802734,29.540499932581923
27.80677032470703,30.42774546184336
28.595569610595703,30.049702960136276
29.388761520385742,30.75252468714298
30.107011795043945,30.67551611865681
30.07628631591797,30.87127830545314
30.238170623779297,30.819989155065176
30.167255401611328,31.044655910894015
30.23528289794922,30.76566807303626
30.142507553100586,30.855270272582906
30.41814422607422,31.085669977142402
30.542020797729492,31.54168848726787
30.552936553955078,31.806797270709346
30.365140914916992,31.745543420280605
30.161026000976562,31.428477390992793
30.008604049682617,31.53076549468323
29.40158462524414,30.849859503848343
29.185991287231445,30.458130804518547
28.994171142578125,31.219558480857543
29.361156463623047,30.801157829568602
29.97194480895996,30.603654271972008
30.55689239501953,30.98374762515382
30.8563232421875,31.272315033110807
30.538822174072266,31.54533619362838
30.777687072753906,31.683189807480737
31.306034088134766,32.04028924199009
31.761159896850586,32.361725973571694
31.952980041503906,32.22233090920504
31.908464431762695,32.18662900901854
31.9650821685791,31.191786264958168
31.710338592529297,31.276372689374966
31.470367431640625,30.697240846225938
31.39305877685547,30.68755908531847
31.224363327026367,29.73798084216071
31.10269546508789,30.125969846700265
30.862136840820312,29.865982585913923
30.824054718017578,30.345339364890854
31.104000091552734,30.71805139586903
31.203495025634766,31.14843115689301
30.844280242919922,32.02970279777044
30.73072624206543,32.219344446542216
30.682048797607422,32.26600725168143
31.365219116210938,32.55287663257064
31.660076141357422,33.09412181382522
31.964412689208984,32.916665278161766
31.57790756225586,32.666524568802274
31.88718032836914,32.77279641930552
32.01460266113281,32.32776636350583
32.557647705078125,32.317406954313896
32.42847442626953,32.246575329187465
32.588531494140625,32.24290261630631
32.2979850769043,32.44431175778065
32.39447784423828,31.92226132502187
32.218536376953125,31.827973507704563
31.975849151611328,31.96444774309644
31.392234802246094,31.808198039341264
31.09642791748047,31.49889658988508
31.328067779541016,31.542126521739007
31.0372257232666,31.57776285868628
30.666175842285156,30.932949675502257
29.892070770263672,31.30926020359346
29.606277465820312,31.448836883531897
29.472763061523438,31.580918935907263
29.44162368774414,31.340230879035737
29.360227584838867,31.850694806507093
29.251745223999023,31.788959504270924
29.4258975982666,31.84357329613966
29.331439971923828,31.733064202776372
28.95069122314453,31.689970060303928
28.62729263305664,31.441092004029475
28.481887817382812,31.05683188476616
28.264602661132812,31.055722655078664
27.883329391479492,29.912019468444164
27.244827270507812,29.488915072063826
27.53708267211914,28.95537385160591
27.063505172729492,28.392213499353588
27.02703094482422,27.217253716232605
26.796634674072266,26.93783220674259
27.18461036682129,25.88731751559067
27.163105010986328,24.628029625824517
26.342260360717773,23.029878691133465
25.59292221069336,21.702322584335306
24.979843139648438,21.103004247903137
24.634078979492188,20.71775233888035
24.302949905395508,20.811932866004206
24.007131576538086,21.419151781089866
23.659475326538086,22.868537986567187
22.890972137451172,23.795470856806283
22.563213348388672,24.693755660097434
22.175792694091797,25.272518774699503
21.978858947753906,25.93622406048769
21.582550048828125,26.51346580411154
21.443592071533203,27.041023796595486
21.737253189086914,26.903696211744776
22.032005310058594,27.438212983307942
22.403053283691406,27.27036932917843
22.599323272705078,27.449671450533124
23.124074935913086,27.332146666829217
23.67070770263672,27.351259950443534
24.136489868164062,27.599762426747066
24.619422912597656,28.002910125860552
24.84868812561035,28.27917021709778
25.421018600463867,28.084117445209237
25.889713287353516,28.000476085156812
26.3126220703125,27.6212997608777
26.47933578491211,27.763368511386204
26.228227615356445,27.710727083382103
26.373233795166016,27.426041037714903
26.26323890686035,27.3775236686588
26.416183471679688,27.461891983198058
25.519176483154297,27.34697667393271
25.08560562133789,27.759404025317586
24.153688430786133,27.934590058063968
24.214488983154297,28.02401598318272
24.138689041137695,27.755689928712446
23.56955337524414,27.52725106221448
22.38479995727539,27.704835712542266
20.630535125732422,27.176573811744326
19.888639450073242,26.59437500298212
19.952037811279297,26.340846256021212
20.300634384155273,26.10941755292401
20.441097259521484,25.736455866781142
20.88311004638672,25.886966705179766
21.287796020507812,25.50137915720287
21.87246322631836,25.8097398392701
22.036409378051758,25.943688911055315
22.870685577392578,26.028085156926824
23.742271423339844,26.167638709877018
24.523544311523438,26.43760844838397
25.139320373535156,27.001031360487122
25.494840621948242,27.310557520837158
26.233745574951172,27.68674345556914
26.88248062133789,27.799435108901275
27.30156135559082,27.832410406809757
27.684249877929688,27.44653838147615
27.69292449951172,27.53284570172521
28.10749053955078,26.805059678837566
27.87183952331543,27.185660895369228
27.788314819335938,26.665792820606548
28.178043365478516,26.68903063555991
28.423171997070312,26.297617543480143
28.24114227294922,26.243294008453326
27.818023681640625,26.36275946219838
27.45240020751953,25.38287334960373
27.649686813354492,25.619787485690505
27.295963287353516,24.370909570172575
27.567970275878906,23.72970532017982
27.16571807861328,22.74799006398417
27.266523361206055,21.930595749065617
26.71919822692871,20.51027618155587
26.362333297729492,19.680237267414476
24.992347717285156,18.971568135119583
24.564197540283203,18.41859688924831
24.398502349853516,18.254083288124402
24.43977165222168,17.697411339303656
23.61229133605957,17.022429288531438
22.93363380432129,16.74020402190533
22.41522789001465,16.774191113241436
}\mytable;

\begin{polaraxis}[
    width=0.8\textwidth,
    height=0.8\textwidth,
    grid=major,
    xtick={-90,-60,-30,0,30,60,90},
    xticklabels={-90,-60,-30,0,30,60,90},
    xmin=-90,
    xmax=90,
    ymin=0,
    ymax=38,
    legend pos=outer north east,
    legend style={draw=black},
    major grid style={gray!30}
]

\addplot[blue, thick] table[
    x expr={\coordindex*180/241 - 90},
    y index=0,
    col sep=comma,
    header=true
] {\mytable};

\addplot[red, thick] table[
    x expr={\coordindex*180/241 - 90},
    y index=1,
    col sep=comma,
    header=true
] {\mytable};

\legend{Original,Revised}
\end{polaraxis}
\end{tikzpicture}}
	}
	\hfill
	\subfloat[\label{fig:secter_pattern_2}]{
		\resizebox{0.149\textwidth}{!}{\begin{tikzpicture}

\pgfplotstableread[col sep=comma]{
23.014080047607422,20.363680780993704
22.56333351135254,19.98851128390642
22.36454200744629,19.219524753827592
22.89422607421875,19.280161949337508
23.370899200439453,18.521223403449838
23.37924575805664,17.777430347239484
23.220176696777344,17.57839977032655
23.435771942138672,17.320647183573563
23.860172271728516,17.50985500953015
23.568513870239258,17.57450744186674
23.86997413635254,16.866355615829207
23.70785140991211,17.417380060811038
24.080089569091797,17.933141298826985
23.99774169921875,17.44674878932852
24.11834716796875,17.471582664474173
23.96782684326172,17.544195296847676
23.889892578125,17.66168679885425
23.833148956298828,17.62232645553845
24.26028823852539,17.998412525429615
24.485857009887695,18.485636295764905
24.847898483276367,18.43258499300158
24.873878479003906,19.17959446451219
25.19087028503418,19.917145829026026
25.710025787353516,20.728123461157885
25.88546371459961,21.906597980490584
25.87705421447754,22.271437217436123
25.31802749633789,23.50923952439051
25.9703426361084,23.954175181754287
26.265132904052734,24.222657924485596
26.489572525024414,25.08523909743481
26.101787567138672,25.657696292914306
25.629331588745117,26.185774682227656
25.97986602783203,26.458412202651317
25.667409896850586,27.122627780039235
26.237905502319336,27.6197943174703
25.89490509033203,28.12867512334629
25.89790153503418,28.31394082927746
25.632232666015625,28.422726703869877
25.77883529663086,28.477753451526276
25.787494659423828,28.562847089437152
25.883079528808594,28.222484073830124
25.643077850341797,27.582058416794077
25.556407928466797,27.335169164474216
24.987276077270508,27.118447674065607
24.712478637695312,26.876842826795006
24.24677085876465,26.635126293878237
23.68830108642578,26.810909241430828
23.47865867614746,26.664626376904373
22.868839263916016,26.200280751565067
22.833311080932617,25.788290624379716
22.006240844726562,25.013959338102612
21.783367156982422,25.065160742690136
21.00104522705078,24.756109226736626
20.70785903930664,24.460404275206177
20.640409469604492,23.974672732456604
20.314369201660156,23.493254020177087
20.352203369140625,22.98068466824469
19.921926498413086,22.750645385274005
20.338455200195312,22.726763275741213
20.42021942138672,22.57499675481916
21.45462417602539,22.913117163222225
22.263891220092773,23.164843207309396
23.261632919311523,23.66312157904645
23.584579467773438,23.762161860838024
24.016555786132812,24.638154055768023
24.039175033569336,24.83846341674603
24.77267074584961,24.360079399056907
25.528846740722656,23.592996967221936
26.81287384033203,23.751986638525295
27.05743408203125,23.302918714444445
28.211191177368164,23.671727920945944
28.646644592285156,24.049671624495836
29.58736801147461,24.957041783290897
29.744766235351562,25.923374512289108
30.352893829345703,26.929605768144814
30.978878021240234,28.430337481353757
31.23163604736328,29.933012587117346
31.902812957763672,30.56367101616784
32.366634368896484,31.62811985578206
32.899391174316406,31.88920752070493
33.1698112487793,32.62277673749273
33.26061248779297,32.77736464810423
33.625755310058594,32.517939991709255
33.76346206665039,32.977099112933026
33.99833679199219,32.58689799988679
34.212127685546875,33.1248124811373
34.505210876464844,33.51591329775896
34.832088470458984,33.88103189774594
34.7416877746582,33.72427962252801
34.8334846496582,33.66869843010424
34.60472869873047,33.5945192551245
34.584327697753906,34.05763441791037
34.200077056884766,33.52366342690393
34.25166702270508,33.51752738293935
34.18598937988281,33.776955620734014
34.23881912231445,33.86260183513022
34.09785461425781,33.99496790017141
34.00017166137695,34.02040820639707
34.02388381958008,33.80818445207129
33.775123596191406,33.90156512193789
33.6303596496582,33.67102112149914
33.432029724121094,33.702225407322565
33.56814956665039,33.51934493849014
34.016395568847656,33.49241997516487
34.42792892456055,33.48086021552075
34.63396453857422,33.14481764053022
34.31937789916992,32.66976286133546
34.46769332885742,32.39405933633689
34.94937515258789,32.37196497507069
35.278892517089844,32.40786223051887
35.41801834106445,32.16482979441176
35.30799102783203,32.2724327906844
35.445648193359375,31.901519808103718
35.50167465209961,31.770361379511087
35.481666564941406,31.688822667275684
35.717811584472656,32.17500753178877
35.86229705810547,31.77658719403516
36.156044006347656,32.83859949596336
36.15208053588867,32.91618215322116
36.07746887207031,33.67647981802602
36.50141525268555,34.06311248116873
36.65497589111328,34.271813831146474
36.62171936035156,34.32394638214478
36.3800048828125,34.35133576651607
36.33251953125,34.40730016188384
36.599361419677734,34.151866453402135
36.64669418334961,34.30114112052032
36.512821197509766,34.08700796111836
35.817466735839844,34.15291830557219
35.83229446411133,34.014204704162964
35.7639274597168,33.74683719364382
35.966087341308594,34.161266080116214
35.600914001464844,33.90454552890199
35.42570114135742,33.55524436400702
35.01084518432617,33.52737350166749
34.754981994628906,32.898153123506205
34.36670684814453,33.19566288758347
34.09354782104492,32.86267694968934
33.43631362915039,32.91537384933012
32.88710021972656,32.67002874315194
32.57905578613281,32.45530864881843
31.94205093383789,31.627932845035023
31.204898834228516,31.03880061508076
30.147682189941406,29.897704420822663
29.492843627929688,29.32930644888092
29.07583236694336,29.117241227577587
28.697345733642578,28.432475218271513
28.24521255493164,28.24145548782444
27.868877410888672,28.128840790569477
27.767398834228516,28.570466952062258
27.699703216552734,28.45926565995591
27.609943389892578,29.32608178331099
27.769977569580078,29.236491281809386
27.92382049560547,28.980904642144026
27.9404239654541,29.057425551688674
27.85015296936035,28.946587961686415
27.673316955566406,28.950249231305204
28.431623458862305,28.30600640838943
28.403644561767578,27.555920071868982
28.639984130859375,27.223480411307342
28.490203857421875,25.85741792417419
28.88787841796875,24.744586153990046
28.955228805541992,23.530705106499678
28.50968360900879,22.428874084333174
28.303665161132812,21.31310974378462
28.263355255126953,19.871833330841394
28.232120513916016,19.764967745092655
27.835182189941406,19.76694749905056
27.32012176513672,19.990180720475337
26.929088592529297,19.80338708100404
26.591279983520508,20.108443788599317
26.62492561340332,20.485222840768497
26.30537986755371,20.755739938264817
25.93025016784668,20.69619706195017
25.374950408935547,20.457502427003035
24.980459213256836,20.41179298537181
24.77798080444336,20.877258005116705
24.287582397460938,20.262488149303604
23.54072380065918,20.335142026925595
22.277469635009766,19.50420182927052
21.151687622070312,19.514352836159016
20.073009490966797,18.66657335169821
19.004690170288086,18.67942642663428
18.3776912689209,18.481154592936587
18.045536041259766,18.47861755305256
18.386066436767578,18.56472532880014
18.817888259887695,19.222944908550343
19.4596004486084,20.03834457786077
20.1488037109375,20.933801418688827
20.710674285888672,21.856302068676033
21.477691650390625,22.553442116230066
22.434402465820312,23.307464435225004
23.742063522338867,23.61421734269524
24.14703369140625,23.758981272917797
24.848102569580078,24.33658436377747
25.08451271057129,23.289028766432942
26.12637710571289,23.535894434161477
26.86984634399414,22.737927286724915
27.086345672607422,22.5745392375456
27.24367332458496,22.5890807802333
27.096820831298828,22.352980113475766
27.542734146118164,22.211018780754426
27.971290588378906,21.810487873790624
28.300365447998047,21.528758428639215
28.241153717041016,22.05866935666498
28.527584075927734,21.597060046910784
28.41192626953125,21.436372606359797
28.662036895751953,21.407697826339692
28.289411544799805,20.948809232911824
28.368450164794922,20.953530021333474
28.207904815673828,20.695690841680477
28.250625610351562,20.955090997906268
28.16575813293457,21.4754165889248
27.92233657836914,21.070218071225113
28.05140495300293,21.469454435292317
28.112625122070312,21.2634637304545
28.01770782470703,21.11402822369695
27.954017639160156,20.847654229230542
27.68842887878418,20.48454406959195
27.922008514404297,20.32182044758756
27.547344207763672,20.429420327409336
27.588821411132812,20.27691771186552
27.99165916442871,20.037200938231475
28.320995330810547,19.83759735809738
28.095489501953125,20.074772170747654
27.795211791992188,19.881572066929767
27.489070892333984,20.190768097397118
27.842243194580078,19.93934714804299
27.711902618408203,19.45734016499688
28.125137329101562,19.549860699178517
27.83216094970703,19.377535630891146
27.985816955566406,19.196536018613905
27.537996292114258,18.86870751967988
27.526762008666992,18.62239494047303
26.618213653564453,18.34324179141475
26.366188049316406,18.466590257083265
26.26323699951172,18.18004270826232
26.450923919677734,18.26239191090735
26.29060935974121,18.01352000794873
26.18366050720215,18.435830924905964
26.182926177978516,18.775588968913823
}\mytable;

\begin{polaraxis}[
    width=0.8\textwidth,
    height=0.8\textwidth,
    grid=major,
    xtick={-90,-60,-30,0,30,60,90},
    xticklabels={-90,-60,-30,0,30,60,90},
    xmin=-90,
    xmax=90,
    ymin=0,
    ymax=38,
    legend pos=outer north east,
    legend style={draw=black},
    major grid style={gray!30}
]

\addplot[blue, thick] table[
    x expr={\coordindex*180/241 - 90},
    y index=0,
    col sep=comma,
    header=true
] {\mytable};

\addplot[red, thick] table[
    x expr={\coordindex*180/241 - 90},
    y index=1,
    col sep=comma,
    header=true
] {\mytable};

\legend{Original,Revised}
\end{polaraxis}
\end{tikzpicture}}
	}
	\caption{The sector antenna array patterns before and after correction of EM algorithm. (a) Sector 0 (b) Sector 23 (c) Sector 27}
	\label{fig:figure_5}
\end{figure}
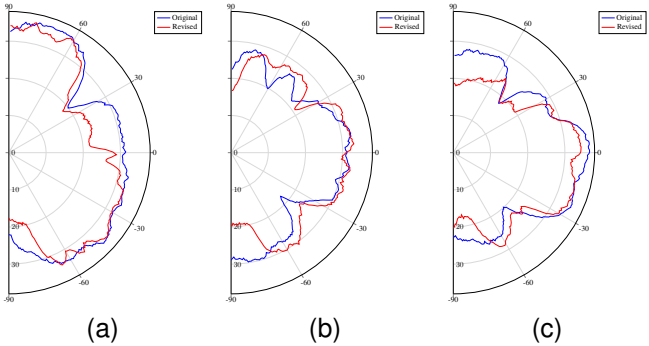

This section examines our algorithm's effectiveness by evaluating the performance of the expectation maximization algorithm and the compression sensing algorithm.

\subsubsection{On estimation of measurement matrix}

We evaluate our enhanced expectation maximization algorithm by analyzing its ability to calibrate antenna patterns. The algorithm is initialized with the measurement matrix $A$ reported in \cite{steinmetzer_compressive_2017}, which represents antenna patterns under ideal conditions. Pattern calibration is performed using BeamSNR measurement frames collected from 40 recording sessions, during which targets execute standardized movement patterns (horizontal left-right and vertical back-forth trajectories). Figure \ref{fig:figure_5} illustrates the transformation from initial to corrected antenna array patterns for three representative sectors (0, 23, and 27). The corrected patterns exhibit significant deviations from the static measurements in \cite{steinmetzer_compressive_2017}, revealing the impact of environmental factors and device-specific variations.

\begin{figure}[htp]
	\centering
	\subfloat[\label{fig:before_em}]{
		\resizebox{0.23\textwidth}{!}{\input{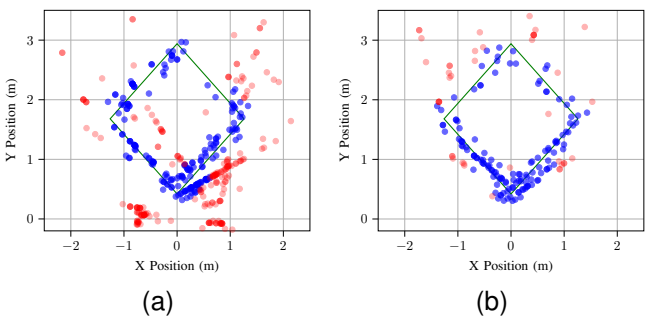}}
	}
	\hfill
	\subfloat[\label{fig:after_em}]{
		\resizebox{0.23\textwidth}{!}{
\begin{tikzpicture}

\definecolor{darkgray176}{RGB}{176,176,176}
\definecolor{green01270}{RGB}{0,127,0}
\definecolor{lightgray204}{RGB}{204,204,204}
\definecolor{steelblue31119180}{RGB}{31,119,180}

\begin{groupplot}[group style={group size=1 by 1}]
\nextgroupplot[
legend cell align={left},
legend style={fill opacity=0.8, draw opacity=1, text opacity=1, draw=lightgray204},
tick align=outside,
tick pos=left,
x grid style={darkgray176},
xlabel={X Position (m)},
xmajorgrids,
xmin=-2.5, xmax=2.5,
xtick style={color=black},
y grid style={darkgray176},
ylabel={Y Position (m)},
ymajorgrids,
ymin=-0.2, ymax=3.5,
ytick style={color=black}
]
\addplot [draw=blue, fill=blue, mark=*, only marks, opacity=0.6]
table{%
x  y
1.16798147702414 1.56934657469598
0.849729741013608 1.27458075553243
1.07752716986482 1.45607777589775
0.808047243304347 1.11393113365783
0.996995505271629 1.24902019779121
0.704724258766738 1.15904799656287
0.867756730356778 1.2889437084261
0.852602409559721 1.31129445807177
0.698935393737263 0.979933783346109
0.64567004851964 1.14197663397733
0.673855139606772 0.833467763884683
0.834498654244309 1.29648136697288
0.809496424884491 0.912866683547681
0.504939668637082 1.02682702664076
0.566134959633349 0.770412704711784
0.703541410571884 1.18932862351981
0.716467979236905 0.858411600285792
0.531236329987898 0.866516728663556
0.294540761958572 0.706436480753703
0.404575933325455 0.780854645339644
0.287560525621989 0.607346370657444
0.381580378730816 0.662381831720984
0.0993014589958808 0.607210949628354
0.225956214574799 0.489020805400667
0.285984306019341 0.740680429243167
0.39163470347874 0.572037034730737
-0.143796828886221 0.433406947622171
0.0796730996469609 0.325045162839826
-0.012294609056632 0.369640805407387
0.141849879040185 0.446877762962172
-0.262706726428463 0.575227246855257
-0.0652784056969001 0.463086045042378
0.185252769028523 0.632523570229311
-0.26747286483458 0.727445626402398
-0.0993015038563586 0.607210949628354
-0.212338403551444 0.66926616501587
-0.362600213309134 0.817555201025233
-0.174507725667056 0.679342040802849
-0.435090464967793 0.894782064392124
-0.287299370087071 0.783194953631143
-0.844180414891003 1.20365835758237
-0.566806727536658 0.994232150963015
-0.694681104821333 1.06157318409698
-0.515274257584152 0.929742361755102
-0.694681104821333 1.06157318409698
-0.515274257584152 0.929742361755102
-0.919501010128363 1.29499765701819
-0.919501010128363 1.29499765701819
-1.2816758325748 1.57592923097576
-0.938844160663442 1.31000174175388
-1.2816758325748 1.57592923097576
-0.979118153026551 1.34124145076429
-1.38914981205624 1.89449085533369
-0.969529357827908 1.52286325414087
-1.31417634551094 1.82809225811426
-1.0083920055053 1.64087486324181
-1.07120749683337 2.03977391096515
-0.574570650313492 2.24204804891407
-0.574570650313492 2.24204804891407
-0.549161926585056 2.56678074847764
-0.261446294911463 2.65149279846598
-0.248283491433696 2.72293019872627
-0.0183930936873846 2.60976678791138
0.0392328242113882 2.8153287425271
0.0918540956287031 2.60689737015102
0.512744953808182 2.27028967912091
0.538378024012888 2.31637557944123
0.651566376369892 2.51987689555669
0.682942482839699 2.57628811355981
0.907168042893217 2.00866221871625
1.0515679554778 1.9150560776118
1.14923559309366 2.01887608805977
1.13558581709438 1.71418737323143
1.26070379756258 1.73455204151181
1.11114561703989 1.6055343043806
0.981585423089682 1.17109366275687
1.03956137017455 1.46426956008926
0.673855139606772 0.833467763884683
0.396758123134762 0.963506841718603
0.510807523685528 0.652424308980023
0.145531246432698 0.752425853026092
0.309538306792162 0.54827443648884
0.375196866411482 0.945391070562178
0.494657589665409 0.644067279680667
0.0492206276064901 0.653944770698865
0.21492462631089 0.499315128752462
0.164643523302157 0.748387146346891
0.317813763661673 0.552556667568236
0.184195405228019 0.667906160378437
0.292853347344874 0.539640484234295
-0.193142160688423 0.398127336603444
0.027970477820906 0.304789354771709
-0.0420526724935734 0.50614932126597
0.179902188106359 0.364312413260019
-0.291149742539336 0.439499145812743
-0.193142160688423 0.398127336603444
-0.0641301505777491 0.343667669634757
0.0390313114282 0.755755470119982
0.146815648051856 0.641181363966003
0.269454867788694 0.510816605774543
-0.240974922286583 0.344260256030819
-0.135839204085936 0.307736577730036
-0.0988153738456283 0.623722173783961
0.0310931931160045 0.528263424714219
0.141849879040185 0.446877762962172
-0.362600213309134 0.817555201025233
-0.193205915416066 0.693081749007284
0.0310931931160045 0.528263424714219
-0.314806066750052 0.782435350578932
-0.174507725667056 0.679342040802849
-0.330335410524634 0.710349620928337
-0.21692944284433 0.902307456087264
-0.50082208880443 1.10776299796018
-0.37121512149328 0.992979445646712
-0.527161206794838 0.990667726703967
-0.407502907849557 0.89785124091759
-0.537235708029828 0.94587996250256
-0.421596126361648 0.860906276261738
-0.825451202599093 1.1576649988368
-0.643027587337444 1.02361741043719
-0.694681104821333 1.06157318409698
-0.706877719468143 0.82752068067124
-0.680441039801379 1.33484157014525
-0.909621109467952 1.21951429951757
-0.643027587337444 1.02361741043719
-1.24443323787067 1.46553923611888
-0.92771668964267 1.23281120123523
-1.03445816862262 1.66519896828008
-1.03445816862262 1.66519896828008
-0.69336217212941 2.17597592769521
-0.297758594626734 2.45441741184955
-0.455007427907533 2.82277924288269
-0.220151839920172 2.87560720648208
0.66657748528123 2.13559614750145
0.66657748528123 2.13559614750145
0.971858223970882 1.40887296699159
1.43062481654389 1.78424876199812
1.22605227644034 1.61686177370701
};
\addplot [draw=red, fill=red, mark=*, only marks, opacity=0.3]
table{%
x  y
0.954808860985211 0.910512648189084
1.15292435253715 1.01632310635607
0.998758746498522 0.933985614679699
0.998758746498522 0.933985614679699
-0.0197053084979996 0.862079038294917
-1.10558827006323 1.05399214207824
-0.88377659013815 0.928000955652903
-1.10558827006323 1.05399214207824
-0.911185530059825 0.94356949947602
-1.35671016539421 1.9659141668686
-1.35671016539421 1.9659141668686
-1.72757257366633 3.1676141668978
-1.12812349314043 2.40120941250637
-1.59995248972061 2.63197454923406
-1.90855077475993 3.58758449306936
-1.15476423051405 2.57038634778848
-1.90855077475993 3.58758449306936
-1.15476423051405 2.57038634778848
-0.524558167076596 2.88122560509095
-0.819405477067539 5.33032158209821
0.349407516987373 3.40444061370061
-2.25530915631104 -4.66136120784565
0.427434124235817 3.08662762010108
0.427434124235817 3.08662762010108
0.460168681764937 3.17244078544019
0.610664198530404 3.5669627297934
1.53353582462797 5.98625850896059
1.21161905720838 3.52679599611687
0.654186754010702 3.68105675868872
0.704123046074854 1.76255036107504
0.914977597880215 0.834267618387846
1.52964849713192 1.96655967176343
0.914977597880215 0.834267618387846
0.902306058151676 0.967193751930099
-0.350762829248162 0.382400083901005
-1.09244283932093 1.01502800759305
-0.879402968268769 0.897661307667827
-1.35671016539421 1.9659141668686
-1.35671016539421 1.9659141668686
-1.68567832289597 3.0318644743613
-1.15509724112247 2.3714119353394
-1.72757257366633 3.1676141668978
-1.16914709656245 2.45365871258257
-0.676634575202582 2.81863106346414
-0.680934627126051 2.65342754073727
-0.586814458732291 3.13154260986393
0.396382207369348 3.00522557464216
0.427434124235817 3.08662762010108
0.427434124235817 3.08662762010108
0.877886966124261 2.45416087044724
1.38713257904548 3.22188640244058
};
    
\addplot [thick, green01270]
table {%
1.26 1.68
0 0.42
-1.26 1.68
0 2.94
1.26 1.68
};
    
\end{groupplot}

\end{tikzpicture}}
	}
	\caption{The localization results with measurement matrix before and after correction. (a) before correction of measurement matrix (b) after correction of measurement matrix}
	\label{fig:figure_6}
\end{figure}

Once extracted through our expectation maximization process, the calibrated measurement matrix is universally applied across all movement patterns in subsequent tests. Figure \ref{fig:figure_6} demonstrates the significant impact of measurement matrix correction on localization performance. Comparing the results before correction (Figure \ref{fig:before_em}) and after correction (Figure \ref{fig:after_em}), we observe two key improvements: a substantial reduction in false detections and an increased density of accurately localized points near the ground-truth trajectory. This enhancement in localization accuracy makes the system more reliable for practical deployments.


\begin{table}[ht!]
	\caption{Impact of EM-based Calibration}
	\vspace{0.5em}\centering
	\begin{tabular}{l cc cc}
		\hline
        State & \multicolumn{2}{c}{Distance Error(m)$\downarrow$} & \multicolumn{2}{c}{Association Rate$\uparrow$} \\
        & Mean & Median & Mean & Median \\
		\hline
        Before Calibration & 0.145 & 0.142 & 0.34 & 0.34 \\
        After Calibration & \textbf{0.138} & \textbf{0.136} & \textbf{0.76} & \textbf{0.76} \\
		\hline
	\end{tabular}
	\label{table_em}
\end{table}

The EM-based calibration results in Table \ref{table_em} demonstrate significant improvements in system performance. The calibrated system achieves a substantially higher association rate compared to pre-calibration, while maintaining strong distance accuracy.

\subsubsection{On compression sensing algorithm}

\begin{figure}[htp]
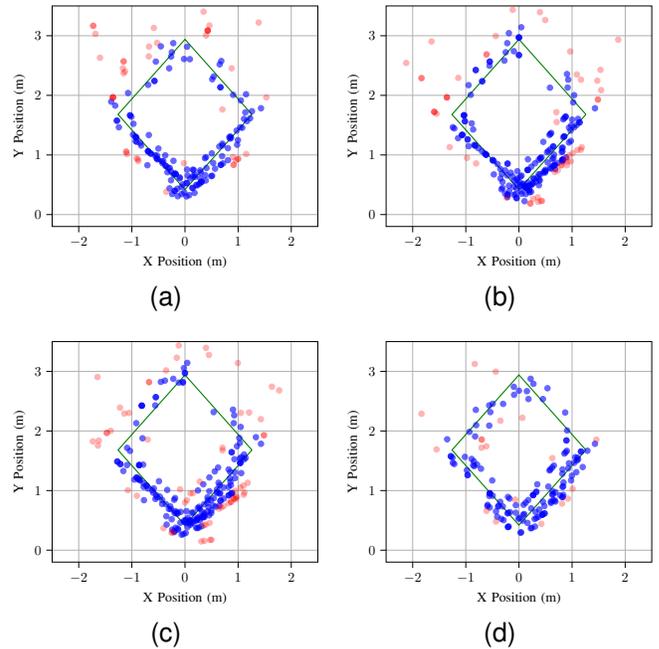

	\centering
	\subfloat[\label{fig:revised_la}]{
		\resizebox{0.23\textwidth}{!}{
\begin{tikzpicture}

\definecolor{darkgray176}{RGB}{176,176,176}
\definecolor{green01270}{RGB}{0,127,0}
\definecolor{lightgray204}{RGB}{204,204,204}
\definecolor{steelblue31119180}{RGB}{31,119,180}

\begin{groupplot}[group style={group size=1 by 1}]
\nextgroupplot[
legend cell align={left},
legend style={fill opacity=0.8, draw opacity=1, text opacity=1, draw=lightgray204},
tick align=outside,
tick pos=left,
x grid style={darkgray176},
xlabel={X Position (m)},
xmajorgrids,
xmin=-2.5, xmax=2.5,
xtick style={color=black},
y grid style={darkgray176},
ylabel={Y Position (m)},
ymajorgrids,
ymin=-0.2, ymax=3.5,
ytick style={color=black}
]
\addplot [draw=blue, fill=blue, mark=*, only marks, opacity=0.6]
table{%
x  y
1.16798147702414 1.56934657469598
0.849729741013608 1.27458075553243
1.07752716986482 1.45607777589775
0.808047243304347 1.11393113365783
0.996995505271629 1.24902019779121
0.704724258766738 1.15904799656287
0.867756730356778 1.2889437084261
0.852602409559721 1.31129445807177
0.698935393737263 0.979933783346109
0.64567004851964 1.14197663397733
0.673855139606772 0.833467763884683
0.834498654244309 1.29648136697288
0.809496424884491 0.912866683547681
0.504939668637082 1.02682702664076
0.566134959633349 0.770412704711784
0.703541410571884 1.18932862351981
0.716467979236905 0.858411600285792
0.531236329987898 0.866516728663556
0.294540761958572 0.706436480753703
0.404575933325455 0.780854645339644
0.287560525621989 0.607346370657444
0.381580378730816 0.662381831720984
0.0993014589958808 0.607210949628354
0.225956214574799 0.489020805400667
0.285984306019341 0.740680429243167
0.39163470347874 0.572037034730737
-0.143796828886221 0.433406947622171
0.0796730996469609 0.325045162839826
-0.012294609056632 0.369640805407387
0.141849879040185 0.446877762962172
-0.262706726428463 0.575227246855257
-0.0652784056969001 0.463086045042378
0.185252769028523 0.632523570229311
-0.26747286483458 0.727445626402398
-0.0993015038563586 0.607210949628354
-0.212338403551444 0.66926616501587
-0.362600213309134 0.817555201025233
-0.174507725667056 0.679342040802849
-0.435090464967793 0.894782064392124
-0.287299370087071 0.783194953631143
-0.844180414891003 1.20365835758237
-0.566806727536658 0.994232150963015
-0.694681104821333 1.06157318409698
-0.515274257584152 0.929742361755102
-0.694681104821333 1.06157318409698
-0.515274257584152 0.929742361755102
-0.919501010128363 1.29499765701819
-0.919501010128363 1.29499765701819
-1.2816758325748 1.57592923097576
-0.938844160663442 1.31000174175388
-1.2816758325748 1.57592923097576
-0.979118153026551 1.34124145076429
-1.38914981205624 1.89449085533369
-0.969529357827908 1.52286325414087
-1.31417634551094 1.82809225811426
-1.0083920055053 1.64087486324181
-1.07120749683337 2.03977391096515
-0.574570650313492 2.24204804891407
-0.574570650313492 2.24204804891407
-0.549161926585056 2.56678074847764
-0.261446294911463 2.65149279846598
-0.248283491433696 2.72293019872627
-0.0183930936873846 2.60976678791138
0.0392328242113882 2.8153287425271
0.0918540956287031 2.60689737015102
0.512744953808182 2.27028967912091
0.538378024012888 2.31637557944123
0.651566376369892 2.51987689555669
0.682942482839699 2.57628811355981
0.907168042893217 2.00866221871625
1.0515679554778 1.9150560776118
1.14923559309366 2.01887608805977
1.13558581709438 1.71418737323143
1.26070379756258 1.73455204151181
1.11114561703989 1.6055343043806
0.981585423089682 1.17109366275687
1.03956137017455 1.46426956008926
0.673855139606772 0.833467763884683
0.396758123134762 0.963506841718603
0.510807523685528 0.652424308980023
0.145531246432698 0.752425853026092
0.309538306792162 0.54827443648884
0.375196866411482 0.945391070562178
0.494657589665409 0.644067279680667
0.0492206276064901 0.653944770698865
0.21492462631089 0.499315128752462
0.164643523302157 0.748387146346891
0.317813763661673 0.552556667568236
0.184195405228019 0.667906160378437
0.292853347344874 0.539640484234295
-0.193142160688423 0.398127336603444
0.027970477820906 0.304789354771709
-0.0420526724935734 0.50614932126597
0.179902188106359 0.364312413260019
-0.291149742539336 0.439499145812743
-0.193142160688423 0.398127336603444
-0.0641301505777491 0.343667669634757
0.0390313114282 0.755755470119982
0.146815648051856 0.641181363966003
0.269454867788694 0.510816605774543
-0.240974922286583 0.344260256030819
-0.135839204085936 0.307736577730036
-0.0988153738456283 0.623722173783961
0.0310931931160045 0.528263424714219
0.141849879040185 0.446877762962172
-0.362600213309134 0.817555201025233
-0.193205915416066 0.693081749007284
0.0310931931160045 0.528263424714219
-0.314806066750052 0.782435350578932
-0.174507725667056 0.679342040802849
-0.330335410524634 0.710349620928337
-0.21692944284433 0.902307456087264
-0.50082208880443 1.10776299796018
-0.37121512149328 0.992979445646712
-0.527161206794838 0.990667726703967
-0.407502907849557 0.89785124091759
-0.537235708029828 0.94587996250256
-0.421596126361648 0.860906276261738
-0.825451202599093 1.1576649988368
-0.643027587337444 1.02361741043719
-0.694681104821333 1.06157318409698
-0.706877719468143 0.82752068067124
-0.680441039801379 1.33484157014525
-0.909621109467952 1.21951429951757
-0.643027587337444 1.02361741043719
-1.24443323787067 1.46553923611888
-0.92771668964267 1.23281120123523
-1.03445816862262 1.66519896828008
-1.03445816862262 1.66519896828008
-0.69336217212941 2.17597592769521
-0.297758594626734 2.45441741184955
-0.455007427907533 2.82277924288269
-0.220151839920172 2.87560720648208
0.66657748528123 2.13559614750145
0.66657748528123 2.13559614750145
0.971858223970882 1.40887296699159
1.43062481654389 1.78424876199812
1.22605227644034 1.61686177370701
};
\addplot [draw=red, fill=red, mark=*, only marks, opacity=0.3]
table{%
x  y
0.954808860985211 0.910512648189084
1.15292435253715 1.01632310635607
0.998758746498522 0.933985614679699
0.998758746498522 0.933985614679699
-0.0197053084979996 0.862079038294917
-1.10558827006323 1.05399214207824
-0.88377659013815 0.928000955652903
-1.10558827006323 1.05399214207824
-0.911185530059825 0.94356949947602
-1.35671016539421 1.9659141668686
-1.35671016539421 1.9659141668686
-1.72757257366633 3.1676141668978
-1.12812349314043 2.40120941250637
-1.59995248972061 2.63197454923406
-1.90855077475993 3.58758449306936
-1.15476423051405 2.57038634778848
-1.90855077475993 3.58758449306936
-1.15476423051405 2.57038634778848
-0.524558167076596 2.88122560509095
-0.819405477067539 5.33032158209821
0.349407516987373 3.40444061370061
-2.25530915631104 -4.66136120784565
0.427434124235817 3.08662762010108
0.427434124235817 3.08662762010108
0.460168681764937 3.17244078544019
0.610664198530404 3.5669627297934
1.53353582462797 5.98625850896059
1.21161905720838 3.52679599611687
0.654186754010702 3.68105675868872
0.704123046074854 1.76255036107504
0.914977597880215 0.834267618387846
1.52964849713192 1.96655967176343
0.914977597880215 0.834267618387846
0.902306058151676 0.967193751930099
-0.350762829248162 0.382400083901005
-1.09244283932093 1.01502800759305
-0.879402968268769 0.897661307667827
-1.35671016539421 1.9659141668686
-1.35671016539421 1.9659141668686
-1.68567832289597 3.0318644743613
-1.15509724112247 2.3714119353394
-1.72757257366633 3.1676141668978
-1.16914709656245 2.45365871258257
-0.676634575202582 2.81863106346414
-0.680934627126051 2.65342754073727
-0.586814458732291 3.13154260986393
0.396382207369348 3.00522557464216
0.427434124235817 3.08662762010108
0.427434124235817 3.08662762010108
0.877886966124261 2.45416087044724
1.38713257904548 3.22188640244058
};
    
\addplot [thick, green01270]
table {%
1.26 1.68
0 0.42
-1.26 1.68
0 2.94
1.26 1.68
};
    
\end{groupplot}

\end{tikzpicture}}
	}
	\hfill
	\subfloat[\label{fig:enet}]{
		\resizebox{0.23\textwidth}{!}{\input{figures/diamond_mmwifi_bpdn.tex}}
	}
	\hfill
	\subfloat[\label{fig:omp}]{
		\resizebox{0.23\textwidth}{!}{\input{figures/diamond_mmwifi_omp.tex}}
	}
	\hfill
	\subfloat[\label{fig:lasso}]{
		\resizebox{0.23\textwidth}{!}{
\begin{tikzpicture}

\definecolor{darkgray176}{RGB}{176,176,176}
\definecolor{green01270}{RGB}{0,127,0}
\definecolor{lightgray204}{RGB}{204,204,204}
\definecolor{steelblue31119180}{RGB}{31,119,180}

\begin{groupplot}[group style={group size=1 by 1}]
\nextgroupplot[
legend cell align={left},
legend style={fill opacity=0.8, draw opacity=1, text opacity=1, draw=lightgray204},
tick align=outside,
tick pos=left,
x grid style={darkgray176},
xlabel={X Position (m)},
xmajorgrids,
xmin=-2.5, xmax=2.5,
xtick style={color=black},
y grid style={darkgray176},
ylabel={Y Position (m)},
ymajorgrids,
ymin=-0.2, ymax=3.5,
ytick style={color=black}
]
\addplot [draw=blue, fill=blue, mark=*, only marks, opacity=0.6]
table{%
x  y
1.08038322806002 1.30863865086939
1.09955247288939 1.55399500900765
0.940927076885262 1.45869378613039
0.820603984475759 1.35489597766915
1.05851898172268 1.56013543303077
0.623396992825895 1.38577344216736
0.673855139606772 0.833467763884683
0.902314254970936 1.66720407448113
0.834749142938469 0.927648672428987
0.575008061922464 1.33694841434154
0.628015357522365 0.931969644580037
0.855587796869702 1.62005670321673
0.819167243032957 1.06124816541859
0.344950566396398 1.10481776389214
0.451970707429369 0.812908324948992
0.590808480756559 1.35289127093278
0.639429037713346 0.939688812615035
0.0799746767198179 0.593393151167915
0.302248244284808 0.396541467961354
0.230513645691117 0.701021476352274
-0.193142160688423 0.398127336603444
0.0425014358446522 0.29865542093074
-0.010366686290481 0.528803271553375
0.222841359774532 0.366616816077027
-0.205969246733288 0.388956566972273
0.0287280657337106 0.293464937668583
0.0102748701982277 0.543561038335404
0.241504507168718 0.373650058564743
-0.205969246733288 0.388956566972273
-0.176505665031944 0.864585286808772
0.0993014589958808 0.607210949628354
-0.408248374905007 0.561640941802694
-0.288473789411283 0.503561593315597
-0.20674171030253 0.892800591753866
-0.0585073686309698 0.754473071862666
-0.303979314332402 0.732998590834679
-0.166084226184832 0.637098316587632
-1.35083698960564 1.86055994906168
-1.01738329284989 1.56524403634692
-0.847137085671929 1.49039695859926
-0.931322598910132 1.98472358084705
-0.889798307288451 1.53020703847142
-0.804604791069501 2.09786327733234
-0.824552421200745 1.90852341260908
-0.811264433688503 2.35372236440202
-0.559937923453552 2.42905950897899
-0.4932952423503 2.45640379958415
-0.0958712092047795 2.74086406414068
-2.37627553190123e-08 2.67537835184368
0.149963653797689 2.66756369798774
0.220151882770452 2.87560720648208
0.451613238398165 2.53584391336636
0.537414531814277 2.71691617667394
0.870370648824766 2.30992215429564
0.907168042893217 2.00866221871625
1.16764305520262 1.65427236580556
1.16764305520262 1.65427236580556
1.20028668820168 1.39436403020663
0.996995505271629 1.24902019779121
0.816805588134758 1.28200445515484
1.11399445176174 1.52517275455214
0.657442757048912 1.42012605925673
0.686672968147132 0.971640525606019
0.878576509168993 1.64325247949834
0.833812496153281 1.07115283349158
0.553482417376845 1.06654597499826
0.603966422663607 0.792557704220286
0.266890848193377 0.924456263524016
0.438728965750554 0.595634456911884
0.175802358975562 0.841647621293939
0.367864603994437 0.560126555878015
0.319781223065735 0.972538867125255
0.477841607268869 0.615232595916674
0.126649667928064 0.796962867525443
0.327652546740566 0.539977633962122
0.33062687442058 0.982398719287615
0.485687130937588 0.619163718942155
0.0876802895738053 0.761535830925904
0.294723166263994 0.52347773503126
0.266890848193377 0.924456263524016
0.438728965750554 0.595634456911884
-0.0593940809314169 0.627830370902254
0.161314783765825 0.456631035932042
0.266890848193377 0.924456263524016
0.438728965750554 0.595634456911884
-0.32948340626552 0.523447379760037
-0.38454227567653 0.74599209430951
-0.204282506767377 0.627466442677797
-0.39236978039295 0.730014188526507
-0.214147358974476 0.616123836573846
-0.549331141215304 0.715820035946117
-0.431204610260405 0.650742447214397
-0.493787533577182 0.913953597062486
-0.352955644841819 0.810468228528036
-0.557817424922591 0.909532394518103
-0.589262465142471 0.760714074999192
-0.502947863066229 1.16921064644349
-0.593595776893882 0.740206054334735
-0.548403302375263 0.954086077089664
-0.548403302375263 0.954086077089664
-0.548403302375263 0.954086077089664
-0.861683331223248 1.42735179905946
-0.965314936782896 1.26043895299298
-1.21668754351986 1.44515128683465
-0.965314936782896 1.26043895299298
-1.25131490564717 1.55237888329843
-0.979118153026551 1.34124145076429
-1.24443323787067 1.46553923611888
-1.2816758325748 1.57592923097576
-1.2816758325748 1.57592923097576
-0.841885439000604 1.40981823242505
-0.78734074395516 1.43459693577746
-0.714430254800483 1.36655921236538
-0.559093597791876 2.08962761499266
-0.538378051489799 2.31637557944123
-0.297758594626734 2.45441741184955
-0.0918541771426959 2.60689737015102
0.179630906682672 2.53567287618022
0.273230523358135 2.79097601990307
0.686170830890443 0.815758967804457
0.894079715935647 1.84138885078522
0.81976670813029 0.89164278033754
0.894079715935647 1.84138885078522
0.81976670813029 0.89164278033754
0.981585423089682 1.17109366275687
1.08834925645921 1.50418920367771
1.43062481654389 1.78424876199812
1.03956137017455 1.46426956008926
0.9299139596871 1.3745529409189
1.28868939690508 1.66811308923642
0.838884789076748 1.13597850014795
0.633601906608089 0.989210764816689
0.640275233358106 0.838303492077888
0.773917455930958 0.91888665180102
0.600010932552337 0.814025055899497
0.31279788162214 0.640842275668171
0.478436922217821 0.740718755441081
0.0739118677909969 0.496799625082697
0.381928593130655 0.682526503608748
};
\addplot [draw=red, fill=red, mark=*, only marks, opacity=0.3]
table{%
x  y
1.01011282348391 1.03029955968582
1.45752300476563 1.85476199905334
0.422550239752137 0.441877466955882
-0.431209500842425 0.480600389492323
-1.83467399022125 2.28905951836358
-0.831320190412796 3.12424659226251
-0.855401107141538 4.0494108382469
-0.43708624819225 2.99426722579165
-0.816074904282501 5.07452952670636
-0.992856628049706 6.21706766042241
1.11114561703989 -1.6055343043806
1.59029426624966 -3.7356711713578
1.03348537334845 -1.53854011371321
1.37499790803158 -3.39200693543884
0.919462218564688 -1.44017694748218
0.858577167108957 -1.38765393633602
0.65515369541293 -1.21216874055205
1.04667804488108 3.79164853684733
0.811209010086692 0.782272424742812
0.920991769057915 0.864680403696063
-0.0687814876367743 0.847961840873917
-0.610594023046683 0.659759448035253
-1.55129490076624 1.69102575454166
-0.705285308990401 1.71789701243879
-0.701690430101699 1.85600826438572
-0.701690430101699 1.85600826438572
-0.613589787300971 1.74337025441799
-0.349675637668173 2.17265161376459
-0.881500536572262 5.0521272848294
0.0644289029911263 2.22144596335377
0.629138540277454 1.54465146587411
};
    
\addplot [thick, green01270]
table {%
1.26 1.68
0 0.42
-1.26 1.68
0 2.94
1.26 1.68
};
    
\end{groupplot}

\end{tikzpicture}}
	}
	\caption{The localization results with different compression sensing algorithms. (a) MMWiLoc (b) enet (c) omp (d) lasso}
	\label{fig:figure_7}
\end{figure}

Figure \ref{fig:figure_7} presents a comprehensive comparison of different compression sensing algorithms with same corrected measurement matrix. Our proposed MMWiLoc algorithm (Figure \ref{fig:revised_la}) demonstrates superior performance compared to ENET (Figure \ref{fig:enet}), OMP (Figure \ref{fig:omp}), and standard LASSO (Figure \ref{fig:lasso}). The enhanced performance of our algorithm can be attributed to its multi-scale nature, which effectively handles varying signal strengths and propagation characteristics across different spatial regions.


\begin{table}[ht!]
	\caption{Compression Sensing Algorithm Comparison}
	\vspace{0.5em}\centering
	\begin{tabular}{l cc cc}
		\hline
        Algorithm & \multicolumn{2}{c}{Distance Error(m)$\downarrow$} & \multicolumn{2}{c}{Association Rate$\uparrow$} \\
        & Mean & Median & Mean & Median \\
		\hline
        MMWiLoc & \textbf{0.135} & \textbf{0.135} & \textbf{0.76} & \textbf{0.76} \\
        LASSO & 0.146 & 0.146 & 0.75 & 0.75 \\
        OMP & 0.147 & 0.147 & 0.66 & 0.66 \\
        ENET & 0.150 & 0.150 & 0.62 & 0.62 \\
		\hline
	\end{tabular}
	\label{table_cs}
\end{table}

The compression sensing algorithm comparison in Table \ref{table_cs} demonstrates the advantages of our multi-scale approach. The MMWiLoc algorithm achieves superior performance in both metrics, with the lowest distance error and highest association rate. This improvement over baseline methods like OMP and ENET indicates that our hierarchical processing structure effectively handles varying signal strengths while maintaining computational efficiency.


These results demonstrate that despite the constraints of millimeter wave Wi-Fi devices, MMWiLoc achieves effective localization through its carefully designed measurement matrix correction and compression sensing algorithms.

\section{Conclusions}
\label{section5}

This paper introduces MMWiLoc, a novel indoor localization system that leverages millimeter wave Wi-Fi signals to achieve centimeter-level accuracy using beamSNR measurements. Our approach combines an enhanced expectation maximization algorithm for measurement matrix refinement with a multi-scale LASSO algorithm for robust target position estimation. Extensive evaluation across diverse movement patterns demonstrates that MMWiLoc achieves mean localization errors as low as 0.12m, significantly outperforming traditional 2.4GHz Wi-Fi methods while maintaining competitive performance with specialized radar systems. The comprehensive multi-sensor dataset provided enables direct performance comparisons across sensing modalities and establishes a valuable benchmark for future indoor localization research. While current limitations include performance degradation at sensing boundaries and single-target operation, MMWiLoc demonstrates the potential of commercial millimeter wave Wi-Fi devices for precise indoor localization with low computational overhead.

\bibliographystyle{IEEEtran}
\bibliography{VariationalPaperCitations.bib}

\begin{thebibliography}{10}
\providecommand{\url}[1]{#1}
\csname url@samestyle\endcsname
\providecommand{\newblock}{\relax}
\providecommand{\bibinfo}[2]{#2}
\providecommand{\BIBentrySTDinterwordspacing}{\spaceskip=0pt\relax}
\providecommand{\BIBentryALTinterwordstretchfactor}{4}
\providecommand{\BIBentryALTinterwordspacing}{\spaceskip=\fontdimen2\font plus
\BIBentryALTinterwordstretchfactor\fontdimen3\font minus \fontdimen4\font\relax}
\providecommand{\BIBforeignlanguage}[2]{{%
\expandafter\ifx\csname l@#1\endcsname\relax
\typeout{** WARNING: IEEEtran.bst: No hyphenation pattern has been}%
\typeout{** loaded for the language `#1'. Using the pattern for}%
\typeout{** the default language instead.}%
\else
\language=\csname l@#1\endcsname
\fi
#2}}
\providecommand{\BIBdecl}{\relax}
\BIBdecl

\bibitem{mautzSurveyOpticalIndoor2011}
R.~Mautz and S.~Tilch, ``Survey of optical indoor positioning systems,'' in \emph{2011 {{International Conference}} on {{Indoor Positioning}} and {{Indoor Navigation}}}, Sep. 2011, pp. 1--7.

\bibitem{gorostizaInfraredSensorSystem2011}
E.~M. Gorostiza, J.~L. L{\'a}zaro~Galilea, F.~J. Meca~Meca, D.~Salido~Monz{\'u}, F.~Espinosa~Zapata, and L.~Pallar{\'e}s~Puerto, ``Infrared {{Sensor System}} for {{Mobile-Robot Positioning}} in {{Intelligent Spaces}},'' \emph{Sensors}, vol.~11, no.~5, pp. 5416--5438, May 2011.

\bibitem{bartolettiSensorRadarNetworks2014}
S.~Bartoletti, A.~Conti, A.~Giorgetti, and M.~Z. Win, ``Sensor {{Radar Networks}} for {{Indoor Tracking}},'' \emph{IEEE Wireless Communications Letters}, vol.~3, no.~2, pp. 157--160, Apr. 2014.

\bibitem{luoVisionTransformersHuman2024}
F.~Luo, S.~Khan, B.~Jiang, and K.~Wu, ``Vision {{Transformers}} for {{Human Activity Recognition Using WiFi Channel State Information}},'' \emph{IEEE Internet of Things Journal}, vol.~11, no.~17, pp. 28\,111--28\,122, Sep. 2024.

\bibitem{liIndoTrackDeviceFreeIndoor2017}
X.~Li, D.~Zhang, Q.~Lv, J.~Xiong, S.~Li, Y.~Zhang, and H.~Mei, ``{{IndoTrack}}: {{Device-Free Indoor Human Tracking}} with {{Commodity Wi-Fi}},'' \emph{Proceedings of the ACM on Interactive, Mobile, Wearable and Ubiquitous Technologies}, vol.~1, no.~3, pp. 1--22, Sep. 2017.

\bibitem{liDynamicMUSICAccurateDevicefree2016}
X.~Li, S.~Li, D.~Zhang, J.~Xiong, Y.~Wang, and H.~Mei, ``Dynamic-{{MUSIC}}: Accurate device-free indoor localization,'' in \emph{Proceedings of the 2016 {{ACM International Joint Conference}} on {{Pervasive}} and {{Ubiquitous Computing}}}.\hskip 1em plus 0.5em minus 0.4em\relax Heidelberg Germany: ACM, Sep. 2016, pp. 196--207.

\bibitem{tanMultiTrackMultiUserTracking2019}
S.~Tan, L.~Zhang, Z.~Wang, and J.~Yang, ``{{MultiTrack}}: {{Multi-User Tracking}} and {{Activity Recognition Using Commodity WiFi}},'' in \emph{Proceedings of the 2019 {{CHI Conference}} on {{Human Factors}} in {{Computing Systems}}}.\hskip 1em plus 0.5em minus 0.4em\relax Glasgow Scotland Uk: ACM, May 2019, pp. 1--12.

\bibitem{qianWidar2PassiveHuman2018}
K.~Qian, C.~Wu, Y.~Zhang, G.~Zhang, Z.~Yang, and Y.~Liu, ``Widar2.0: {{Passive Human Tracking}} with a {{Single Wi-Fi Link}},'' in \emph{Proceedings of the 16th {{Annual International Conference}} on {{Mobile Systems}}, {{Applications}}, and {{Services}}}.\hskip 1em plus 0.5em minus 0.4em\relax Munich Germany: ACM, Jun. 2018, pp. 350--361.

\bibitem{nitsche_ieee_2014}
T.~Nitsche, C.~Cordeiro, A.~B. Flores, E.~W. Knightly, E.~Perahia, and J.~C. Widmer, ``{IEEE} 802.11ad: directional 60 {GHz} communication for multi-gigabit-per-second wi-fi [invited paper],'' \emph{{IEEE} Communications Magazine}, vol.~52, no.~12, pp. 132--141, 2014.

\bibitem{steinmetzer_compressive_2017}
D.~Steinmetzer, D.~Wegemer, M.~Schulz, J.~Widmer, and M.~Hollick, ``Compressive millimeter-wave sector selection in off-the-shelf {IEEE} 802.11ad devices,'' in \emph{Proceedings of the 13th International Conference on emerging Networking {EXperiments} and Technologies}.\hskip 1em plus 0.5em minus 0.4em\relax {ACM}, 2017, pp. 414--425.

\bibitem{bielsa_indoor_2018}
G.~Bielsa, J.~Palacios, A.~Loch, D.~Steinmetzer, P.~Casari, and J.~Widmer, ``Indoor localization using commercial off-the-shelf 60 {GHz} access points,'' in \emph{{IEEE} {INFOCOM} 2018 - {IEEE} Conference on Computer Communications}, 2018, pp. 2384--2392.

\bibitem{vaca-rubio_object_2024}
C.~J. Vaca-Rubio, P.~Wang, T.~Koike-Akino, Y.~Wang, P.~Boufounos, and P.~Popovski, ``Object trajectory estimation with continuous-time neural dynamic learning of millimeter-wave wi-fi,'' \emph{{IEEE} Journal of Selected Topics in Signal Processing}, vol.~18, no.~5, pp. 796--811, 2024.

\bibitem{yang_mm-fi_2023}
J.~Yang, H.~Huang, Y.~Zhou, X.~Chen, Y.~Xu, S.~Yuan, H.~Zou, C.~X. Lu, and L.~Xie, ``{MM}-fi: Multi-modal non-intrusive 4d human dataset for versatile wireless sensing.''

\bibitem{jiang_towards_2020}
W.~Jiang, H.~Xue, C.~Miao, S.~Wang, S.~Lin, C.~Tian, S.~Murali, H.~Hu, Z.~Sun, and L.~Su, ``Towards 3d human pose construction using wifi,'' in \emph{Proceedings of the 26th Annual International Conference on Mobile Computing and Networking}, ser. {MobiCom} '20.\hskip 1em plus 0.5em minus 0.4em\relax Association for Computing Machinery, 2020, pp. 1--14.

\bibitem{liao_wisense_2024}
Z.~Liao, J.~Su, Y.~Ye, and R.~Q. Hu, ``Wisense: A dataset for {WiFi}-based human activity recognition,'' \emph{{IEEE} Wireless Communications}, vol.~31, no.~5, pp. 232--237, 2024.

\bibitem{thanh_public_2024}
P.-V.~L. Thanh, Q.~N. Ph., D.~Nguyen, B.~Khuc, A.~Gelgor, and P.~N. T.H., ``Public dataset for simultaneous human activity recognition and localization using {WiFi} signals,'' in \emph{2024 International Conference on Electrical Engineering and Photonics ({EExPolytech})}, 2024, pp. 206--209.

\bibitem{huang_wimans_2024}
S.~Huang, K.~Li, D.~You, Y.~Chen, A.~Lin, S.~Liu, X.~Li, and J.~A. {McCann}, ``{WiMANS}: A benchmark dataset for {WiFi}-based multi-user activity sensing,'' in \emph{Computer Vision – {ECCV} 2024: 18th European Conference, Milan, Italy, September 29–October 4, 2024, Proceedings, Part {XLII}}.\hskip 1em plus 0.5em minus 0.4em\relax Springer-Verlag, 2024, pp. 72--91.

\bibitem{yang_sensefi_2023}
J.~Yang, X.~Chen, D.~Wang, H.~Zou, C.~X. Lu, S.~Sun, and L.~Xie, ``{SenseFi}: A library and benchmark on deep-learning-empowered {WiFi} human sensing.''

\bibitem{meneghello_csi_2023}
F.~Meneghello, N.~D. Fabbro, D.~Garlisi, I.~Tinnirello, and M.~Rossi, ``A {CSI} dataset for wireless human sensing on 80 {MHz} wi-fi channels,'' \emph{{IEEE} Communications Magazine}, vol.~61, no.~9, pp. 146--152, 2023.

\bibitem{7znf-qp86-20}
Z.~Yang, Y.~Zhang, G.~Zhang, Y.~Zheng, and G.~Chi, ``Widar 3.0: Wifi-based activity recognition dataset,'' 2020.

\bibitem{guo_huac_2018}
L.~Guo, L.~Wang, J.~Liu, W.~Zhou, and B.~Lu, ``{HuAc}: Human activity recognition using crowdsourced {WiFi} signals and skeleton data,'' \emph{Wireless Communications and Mobile Computing}, vol. 2018, no.~1, p. 6163475, 2018.

\bibitem{yousefi_survey_2017}
S.~Yousefi, H.~Narui, S.~Dayal, S.~Ermon, and S.~Valaee, ``A survey on behavior recognition using {WiFi} channel state information,'' \emph{{IEEE} Communications Magazine}, vol.~55, no.~10, pp. 98--104, 2017.

\bibitem{ma_signfi_2018}
Y.~Ma, G.~Zhou, S.~Wang, H.~Zhao, and W.~Jung, ``{SignFi}: Sign language recognition using {WiFi},'' \emph{Proc. {ACM} Interact. Mob. Wearable Ubiquitous Technol.}, vol.~2, no.~1, pp. 23:1--23:21, 2018.

\bibitem{zou_regularization_2005}
H.~Zou and T.~Hastie, ``Regularization and variable selection via the elastic net,'' \emph{Journal of the Royal Statistical Society Series B: Statistical Methodology}, vol.~67, no.~2, pp. 301--320, 2005.

\bibitem{cai_orthogonal_2011}
T.~T. Cai and L.~Wang, ``Orthogonal matching pursuit for sparse signal recovery with noise,'' \emph{{IEEE} Transactions on Information Theory}, vol.~57, no.~7, pp. 4680--4688, 2011.

\bibitem{tibshirani_regression_1996}
R.~Tibshirani, ``Regression shrinkage and selection via the lasso,'' \emph{Journal of the Royal Statistical Society: Series B (Methodological)}, vol.~58, no.~1, pp. 267--288, 1996.

\bibitem{wuWiTrajRobustIndoor2021}
D.~Wu, Y.~Zeng, R.~Gao, S.~Li, Y.~Li, R.~C. Shah, H.~Lu, and D.~Zhang, ``{{WiTraj}}: {{Robust Indoor Motion Tracking}} with {{WiFi Signals}},'' \emph{IEEE Transactions on Mobile Computing}, pp. 1--1, 2021.

\end{thebibliography}


 




\vfill

\end{document}